\documentclass[11pt, a4paper]{article}   	
\pdfoutput=1
\usepackage{jheppub}
\usepackage{graphicx}
\usepackage{graphicx}				
\usepackage{xcolor}								
\usepackage{amssymb}
\usepackage{amsmath}
\usepackage{bm}

\newcommand{\be}{\begin{equation}}
\newcommand{\ee}{\end{equation}}

\newcommand{\bra}[1]{\left\langle #1 \right|}
\newcommand{\ket}[1]{\left| #1 \right\rangle}
\newcommand{\braket}[3]{\left \langle #1 \left | #2 \right| #3 \right\rangle}
\newcommand{\brkt}[2]{\left \langle #1| #2 \right\rangle}
\newcommand{\avg}[1]{\left\langle #1\right\rangle}

\newcommand{\tr}{\text{tr }}

\newcommand{\GCD}{\text{GCD}}

\title{Lattice Quantum Villain Hamiltonians: Compact scalars, $U(1)$ gauge theories, fracton models and Quantum Ising model dualities}
\author{Lucca Fazza, Tin Sulejmanpasic}
\affiliation{Department of Mathematical Sciences, Durham University, South Road, DH1 3LE Durham, United Kingdom}
\emailAdd{lucca.fazza-marcon@durham.ac.uk}
\emailAdd{tin.sulejmanpasic@durham.ac.uk}
\abstract{We construct Villain Hamiltonians for compact scalars and abelian gauge theories. The Villain integers are promoted to integral spectrum operators, whose canonical conjugates are naturally compact scalars. Further, depending on the theory, these conjugate operators can be interpreted as (higher-form) gauge fields. If a gauge symmetry is imposed on these dual gauge fields, a natural constraint on the Villain operator leads to the absence of defects (e.g. vortices, monopoles,...). These lattice models therefore have the same symmetry and anomaly structure as their corresponding continuum models. Moreover they can be formulated in a way that makes the well-know dualities look manifest, e.g. a compact scalar in 2d has a T-duality, in 3d is dual to a U(1) gauge theory, etc. We further discuss the gauged version of compact scalars on the lattice, its anomalies and solution, as well as a particular limit of the gauged XY model at strong coupling which reduces to the transverse-field Ising model. The construction for higher-form gauge theories is similar. We apply these ideas to the constructions of some models which are of interest to fracton physics, in particular the XY-plaquette model and the tensor gauge field model. The XY-plaquette model in 2+1d coupled to a tensor gauge fields at strong gauge coupling is also exactly described by a transverse field quantum $J_1-J_2$ Ising model with  $J_1=2J_2$, and discuss the phase structure of such models. }

\begin{document}
\maketitle

\section{Introduction}

Naive lattice discretization of quantum field theories can lead to a reduced symmetry group. This is especially true if the symmetries in question have a mixed 't Hooft anomalies. The most familiar example is that of a massless free Dirac fermion in 2d and 4d, in which case the symmetry group is $U(1)_V\times U(1)_A$ where the index stands for \emph{vector} and \emph{axial}.  The two $U(1)$ symmetries famously have a mixed triangle anomaly, as well as a mixed axial--gravitational anomaly, and the lattice discretization was for a long time taught to be impossible preserving the axial symmetry. Yet L\"uscher \cite{Luscher:1998pqa}, building on the works of Ginsparg and Wilson \cite{Ginsparg:1981bj} as well as Neuberger \cite{Neuberger:1997fp}, constructed such a lattice action with the correct anomaly.

A closely related example is a compact scalar in 2d, which is a bosonized version of a 2d Dirac fermion. The usual way to discretize the compact boson is by an XY-model, but this model has a reduced symmetry group. Namely the winding symmetry, under which the winding charge $Q=\frac{1}{2}\int dx \partial_x\phi$ of the compact scalar $\phi$  is not conserved, because the lattice theory contains dynamical vortices which can induce the famous Kosterlitz-Thouless-Berezinskii transition. Another example are $U(1)$ abelian gauge theories in 3 space-time dimensions and higher, whose naive lattice discretization has dynamical monopoles which violate a monopole symmetry. Such theories were discretized using Modified Villain Actions in \cite{Sulejmanpasic:2019ytl}, in which a famous Villain model was modified to incorporate a no-defect (i.e. no-vortex or no-monopole) constraint, and hence enhance the global symmetries. Such models were applied to fracton models in \cite{Gorantla:2021svj} and for constructing non-invertible symmetries in \cite{Choi:2021kmx}.

In this paper we show that such $U(1)$ theories have a natural Hamiltonian formulation which we dub Villain Hamiltonians\footnote{After this draft was largely finished we found out that the upcoming publication \cite{Cheng:2022sgb} which has a discussion on the Hamiltonian formulation of compact scalars. See also the discussion in \cite{Yoneda:2022qpj} from a different perspective, for some compact scalar models.}. The idea is to introduce integer-spectrum operators -- the Villain operators -- which have a natural angle-valued (i.e. circle-valued) operator as its canonical conjugate. Depending on the theory, the conjugate operator can be interpreted as gauge field, and by imposing a gauge symmetry, a form of Gauss law constrains the Villain operator, which exactly implements the no-defect constraint.

\section{Compact scalar in 1 spatial dimension}

Consider a natural lattice discretization Hamiltonian of a free massless discrete scalar theory
\be
H=\sum_x\Big(\frac{1}{2Ja}\pi_x^2+ \frac{J}{2a}(\phi_{x+1}-\phi_x)^2\Big)\;,
\ee
with $[\phi_x,\pi_y]=i\delta_{xy}$, and where $J$ is dimensionless, while $a$ has dimensions of length. The constant $a$ above is the only constant with dimension which sets the scale of the problem. 
We want to promote $\phi_x$ to be a compact scalar, i.e. that $\phi_x\sim \phi_x+2\pi$. This is impossible with the Hamiltonian above, as shift of $\phi_x$ by $2\pi$ on distinct sites is not a symmetry. Instead we go to a Villain-type Hamiltonian
\be\label{eq:H1p1d}
H=\sum_x\Big(\frac{1}{2Ja}\pi_x^2+ \frac{J}{2a}(\phi_{x+1}-\phi_x+2\pi n_{x})^2\Big)\;,
\ee
where $n_x$ is an operator with only integer eigenvalues. To such an operator one naturally associates an angle-valued operator $\tilde\phi_x$, with canonical commutation relations
\be
[\tilde \phi_x,n_y]=i\delta_{xy}\;.
\ee
Further we will assume that $[n_x,\phi_y]=[n_x,\pi_y]=[\tilde\phi_x,\phi_y]=0$.
The above implies that $e^{i2\pi n_x}\tilde\phi_x e^{-i2\pi n_x}=\tilde\phi_x+2\pi$. Since this shift is supposed to be a gauge symmetry, the Hilbert space is invariant, and hence $e^{i2\pi n_x}\ket{\Psi}=\ket\Psi$, and so $n_x$ can take only integer values. For this to be self-consistent we need also to have that $e^{2\pi i n_x}$ commutes with the Hamitlonian, which is true by our assumption that $\tilde\phi_x$ commutes with both $\pi_x$ and $\phi_x$.

Now let us look for a transformation which shifts $\phi_x$ by $2\pi k_x$ where $k_x$ are integers, in such a way that it is an invariance of the Hamiltonian. The naive transformation $e^{i2\pi \sum_x k_x\pi_x}$ does not do the job, as the Hamiltonian is not invariant under it. Indeed
\be
e^{i2\pi \pi_x}He^{-i2\pi \pi_x}= \sum_x \left( \frac{1}{2Ja}\pi_x^2+\frac{J}{2a}(\phi_{x+1}-\phi_x+2\pi (k_{x+1}-k_x+n_x))^2\right)\;.
\ee
But now we want to shift $n_x\rightarrow n_x-(k_{x+1}-k_x)$. To do that we use the operator
\be
e^{i\sum_x\tilde\phi_x (k_{x+1}-k_x)}\;.
\ee
The total operator which implements the shift of $\phi_x$ by $2\pi k_x$ and is an invariance of the Hamiltonian is
\be
e^{i2\pi \sum_x\pi_x}e^{i\sum_x\tilde\phi_x (k_{x+1}-k_x)}\;.
\ee
Note that $k_x$ can be arbitrary integers. Now  since $\pi_x$ and $\tilde\phi_y$ commute for any $x,y$, the operator which shifts $\phi_x$ by $2\pi k_x$ is given by
\be\label{eq:constraint}
e^{i\sum_x k_x(2\pi\pi_x+\tilde\phi_{x-1}-\tilde\phi_x)}=\mathbb I\;, \qquad k_x\in\mathbb Z
\ee
where we demand that the operator must be acting trivially on the Hilbert space for any $k_x\in \mathbb Z$, which implies
\be
2\pi \pi_x+\tilde\phi_{x-1}-\tilde\phi_x= 2\pi \tilde n_x\;,
\ee
where $\tilde n_x$ is some integer-valued operator. Expressing $\pi_x$ in terms of $\tilde n_x$ we have
\be
\pi_x= \frac{1}{2\pi}\left(\tilde \phi_{x}-\tilde\phi_{x-1}+2\pi \tilde n_x\right)\;.
\ee
To keep the canonical commutation relations $[\phi_x,\pi_y]=i\delta_{xy}$ we impose the relation $[\phi_x,\tilde n_x]=i\delta_{xy}$. We also demand $[\pi_x,n_y]=0$ for all $x,y$ and so
\be
[\pi_x,n_y]=\frac{1}{2\pi}[\tilde \phi_x-\tilde\phi_{x-1}+2\pi \tilde n_x,n_y]=\frac{i\delta_{x,y}-i\delta_{x-1,y}}{2\pi}+[\tilde n_x, n_y]=0
\ee
so that
\be
[n_x,\tilde n_y]=i \frac{\delta_{x,y}-\delta_{x,y-1}}{2\pi}\;.
\ee

Expressing the Hamitonian now yields
\be
H= \sum_x \Bigg[\frac{1}{2Ja(2\pi)^2}\left(\tilde \phi_x-\tilde\phi_{x-1}+2\pi \tilde n_x\right)^2+\frac{J}{2a}\left(\phi_{x+1}-\phi_x+2\pi n_x\right)^2\Bigg]\;,
\ee
with the following commutation relations
\begin{align}
&[\phi_x,\tilde n_y]=i\delta_{xy}\;,&&[\tilde\phi_x,n_y]=i\delta_{xy}\;.\\
&[n_x,\tilde n_y]= i\frac{\delta_{x,y}-\delta_{x,y-1}}{2\pi}
\end{align}

The above Hamiltonian and the canonical commutation relations are invariant under the change 
\begin{align}
&J\rightarrow \frac{1}{(2\pi)^2J}\\
&\tilde \phi_x\rightarrow \phi_{x+1}\;,&&\phi_x\rightarrow \tilde\phi_x\;,\\
&\tilde n_x\rightarrow n_x\;,&&n_x\rightarrow \tilde n_{x+1}\;.
\end{align}
This is the self-duality transformation. Note that the self-duality becomes a symmetry when $J=1/(2\pi)$. However, rather than squaring to identity, it squares to a lattice translation. So, self-duality  at the special point is an extension of the translation symmetry. 

Further, the spectrum of the above Hamiltonian can be solved exactly, as we show in the Appendix \ref{app:compact_boson}. By expanding $\phi_x$ and $n_x$ into Fourier modes, we get that the Hamiltinian reduces to

\begin{equation}
	H=\frac{J(2\pi)^2}{2a}N\left(\frac{\tilde\Pi}{N}\right)^2+\frac{N}{2Ja}\left(\frac{\Pi}{N}\right)^2+\sum_{p}\omega_{p}\left(B_pB_p^\dagger+B_p^\dagger B_p\right)
\end{equation}
where the sum over $p$ is over $p=\frac{2\pi}{N},\frac{4\pi}{N},\dots, \frac{2\pi(N-1)}{N}$, $\Pi$ is the conserved charge due to the global shift symmetry $\phi_x\rightarrow \phi_x+\text{constant}$, $\tilde\Pi$ is the charge due to the global shift symmetry $\tilde\phi_{\tilde x}+\text{constant}$, the operators $B_p$ and $B_p^\dagger$ (defined only for $p\ne 0 \bmod 2\pi$) satisfy the commutation relation $[B_{p},B_{p'}^\dagger]=\delta_{p,p'}$, with the dispersion relation being
\be
\omega_p^2=\frac{4\sin^2\left(\frac{p}{2}\right)}{a^2}\;.
\ee
The exact solution is a direct lattice analogue of the continuum compact scalar theory. The nontrivial fact is that the zeromode contributions containing $\Pi$ and $\tilde \Pi$ appear naturally. 

We can look at the spatial correlator \eqref{eq:correlator_result}
\be
\avg{:e^{i\phi_x}::e^{-i\phi_y}:}=e^{-\frac{1}{2JNa}\sum_{p\ne 0} \frac{e^{ip(x-y)}}{\omega_p}}\;,
\ee
where $::$ indicates normal ordering of $B_p$ and $B_p^\dagger$ operators. Now it is natural to interpret $a$ as the UV lattice size and take a continuum limit to be $N\rightarrow\infty$ and $a\rightarrow 0$ such that $L=Na$ is fixed. Then we define the dimensionful coordinate held fixed in the continuum limit as $x_c=xa$ and obtain that
\be\label{eq:correlator_cont}
\avg{:e^{i\phi_0}::e^{-i\phi_{x_ca}}:}\rightarrow e^{-\frac{1}{2J}\sum_{p=1}^\infty \frac{e^{i\frac{2\pi p x_c}{L}}}{{2\pi |p|}}}\;,
\ee
which is the correct continuum finite-volume expression for the correlator\footnote{The continuum Lagrangian is $L=\frac{J}{2}(\partial_\mu\phi)^2$, and the spatial correlator at finite volume $L$ is given by $\avg{:e^{i\phi(x)}::e^{-i\phi(y)}:}=\exp\left(-\frac{2}{J}\frac{1}{L}\sum_{p=1}^\infty\int \frac{dk_0}{2\pi}\frac{e^{i\frac{2\pi p(x-y)}{L}}}{k_0^2+\left(\frac{2\pi p}{L}\right)^2}\right)$, which upon integration over $k_0$, is equal to \eqref{eq:correlator_cont}.}. 

\subsection{Going to a space-time lattice}

Now consider the Hamiltonian \eqref{eq:H1p1d}, and let us construct the space-time lattice by writing
\be
Z=\tr \underbrace{e^{-\epsilon \hat H}\times e^{-\epsilon \hat H}\times\cdots e^{-\epsilon \hat H}}_{\text{$N_t$ times}}\;.
\ee
We now want to insert complete sets of states. Since\footnote{We insert hats for operators in this section to distinguish from their eigenvalues which are without hats.} $\hat{\tilde \phi}_x$ and $\hat \pi_y$ commute for all $x,y$, we can construct simultaneous eigenstates $\ket{\{\tilde \phi\},\{\pi\}}$. Similarily we can do the same for $\ket{\{n\},\{\phi\}}$ The inner product between the two is given by
\be
\brkt{\{\tilde \phi\},\{\pi\}}{\{n\},\{\phi\}}=\frac{1}{2\pi}{e^{i\sum_x n_x\tilde\phi_x-i\sum_x\pi_x\phi_x}}\;.
\ee
We write
\be
e^{-\epsilon H}\approx \prod_{y}e^{- \frac{\epsilon}{2J}\hat\pi_y^2}\prod_{y}e^{-\frac{J\epsilon}{2}(\hat\phi_{y+1}-\hat\phi_y+2\pi \hat n_y)^2}\;,
\ee
which is valid for sufficiently small $\epsilon$. Now we insert complete sets of states and obtain 
\begin{multline}
	Z\approx \int D\Phi\bra{\{\tilde \phi^0\},\{\pi^0\}}e^{- \frac{\epsilon}{2J}\hat\pi_x^2}\ket{\{n^0\},\{\phi^0\}}\bra{\{n^0\},\{\phi^0\}}e^{-\frac{J\epsilon}{2a}(\hat\phi_{x+1}-\hat\phi_x+2\pi \hat n_x)^2}\\\ket{\{\tilde\phi^1\},\{\pi^1\}}\bra{\{\tilde\phi^1\},\{\pi^1\}}e^{- \frac{\epsilon}{2Ja}\hat\pi_x^2}\ket{\{n^1\},\{\phi^1\}}\bra{\{n^1\},\{\phi^1\}}e^{-\frac{J\epsilon}{2a}(\hat\phi_{x+1}-\hat\phi_x+2\pi \hat n_x)^2}\\\cdots e^{- \frac{\epsilon}{2Ja}\hat\pi_x^2}\ket{\{n^{N_t-1}\},\{\phi^{N_t-1}\}}\bra{\{n^{N_t-1}\},\{\phi^{N_t-1}\}}e^{-\frac{J\epsilon}{2a}(\hat\phi_{x+1}-\hat\phi_x+2\pi \hat n_x)^2}\ket{\tilde \phi^{N_t}_x,\pi_x^{N_t}}=\\=\int D\Phi e^{-\sum_{x,t}\left[\frac{\epsilon}{2Ja}(\pi_x^t)^2+\frac{J\epsilon}{2a}(\phi_{x+1}^t-\phi_x^t+2\pi n^t_x)-in_x^t(\tilde \phi_x^{t+1}-\tilde \phi_x^t)+i\phi_x^t(\pi_x^{t+1}-\pi_x^{t})\right]}
	\;.
\end{multline}
The measure $D\Phi$ is just a yet unspecified integration measure over $\phi_x^t,\tilde \phi_x^t,\pi_x^t$ and $n_x^t$ which we will fix in a moment. Note that the sum over $x$ and $t$ runs from $x=0,\dots N_t-1$ and $t=0,\dots, N-1$ where we identify variables at $x=0$ and $x=N$ and at $t=0$ and $t=N_t$. 

Now to specify the integration measure we have to remember to implement the constraint that 
\be
e^{ \sum_x i k_x \left(2\pi\hat\pi_x+{\hat{\tilde \phi}_{x-1}-\hat{\tilde\phi}_{x}}\right)}=\mathbb I\;.
\ee
To do that we pick an integration measure
\be
\int d\Phi= \sum_{\{n,k\}}\int D\phi \int D\tilde\phi \int D\pi \;e^{ i\sum_x k^t_x \left(2\pi\pi^t_x+{{\tilde \phi}^t_{x}-{\tilde\phi}^t_{x-1}}\right)}\;,
\ee
where the sum over integers $k_x^t$ implements the appropriate constraint.
The expression for the partition function is then
\begin{multline}
	Z\approx \sum_{\{n,k\}}\int D\phi \int D\tilde\phi \int D\pi\; \\e^{-\sum_{x,t}\left[\frac{\epsilon}{2Ja}(\pi_x^t)^2+\frac{J\epsilon}{2a}(\phi_{x+1}^t-\phi_x^t+2\pi n^t_x)-in_x^t(\tilde \phi_x^{t+1}-\tilde \phi_x^t)-i\pi_x^{t}(\phi_x^{t}-\phi_x^{t-1}+2\pi k_x^t)-ik_x^t(\tilde\phi_x^t-\tilde \phi_{x-1}^{t})\right]}\;.
\end{multline}
Integrating over $\int D\pi$ yields
\begin{multline}
Z\approx \sum_{n,k}\int D\phi \int D\tilde\phi \\\times e^{-\sum_{x,t}\Big(\frac{Ja}{2\epsilon}(\phi_x^t-\phi_x^{t-1}+2\pi k_x^t)^2+\frac{J\epsilon}{2a}(\phi_{x+1}^t-\phi_x^{t}+2\pi n_x^t)^2-i n_x^t(\tilde \phi_x^{t+1}-\tilde\phi^t_x)-ik_x^t(\tilde\phi_x^t-\tilde\phi_{x-1}^t)\Big)}\;.
\end{multline}

Now if we set $a=\epsilon$ and we relabel $(n_{(x,t),1},n_{(x,t),2})=(k_{x-1}^{t},n_x^t,)$, we can write the above action more concisely as
\be
Z\approx \sum_{n,k}\int D\phi \int D\tilde\phi e^{-\sum_{\bm x}\sum_{\mu=1,2}\Big(\frac{J}{2}(\phi_{\bm x+\hat\mu}^t-\phi_{\bm x}+2\pi n_{\bm x,\mu})^2-i n_{\bm x,\mu}(\tilde \phi_{\bm x+\hat\mu}-\tilde\phi_{\bm x})\Big)}\;.
\ee
which is just the modified Villain formulation \cite{Sulejmanpasic:2019ytl,Gorantla:2021svj}.

\section{The U(1) gauge theories}

Here we will discuss U(1) gauge theories. We will start by discussing the ordinary (i.e. 1-form gauge theories) in 2 and 3 spatial dimensions. Then we will discuss a general $p$-form $U(1)$ gauge field in arbitrary number of dimensions. 

We find it convenient to introduce co-chain notation, which we review here. Our notation will follow that of the appendix of \cite{Sulejmanpasic:2019ytl}. A lattice\footnote{Most of what we say here applies for any graph without any special symmetry properties.} $\Lambda$ in arbitrary number of dimensions $D$ has sites, which we will label with $x$ or $y$ (0-cells), links $l$ (1-cells), plaquettes $p$ (2-cells), cubes $c$ (3-cells), hypercubes $h$ (4-cells) or in general $r$-cells $c^r$. Since we discuss Hamiltonians in this work, our lattice is a spatial lattice only. $r$-cells of a the lattice can be formally added together with arbitrary coefficients (which are typically taken to be integers) to form an  r-chain. The lattice is sometimes referred to as a \emph{cell-complex} or \emph{CW complex} in the math literature. An $r$-chain then forms a group $C_r(X)$, where $X$ is the manifold on which the lattice lives. Operators such as the boundary operator $\partial$ maps an $r$-cell $c^r$ into a linear combination of $(r-1)$-cells -- the boundary cells of the $c^r$. Note that $r$-cells have an orientation. Two $r$-cells which are the same, but have a different orientation are taken to formally differ by a sign in front. The orientation of the $(r-1)$-cells in $\partial c^{r}$ is taken to be outward. 

We can define a dual lattice $\tilde \Lambda$. Sites $\tilde x$ associated with the dual lattice are $D$-cells of the original lattice, links of the dual lattice are $D-1$ cells of the original lattice, and so on. An $r$-cell $c^r$ of the lattice $\Lambda$ intersects an $D-r$ cell of the dual-lattice. Therefore there is a natural map from $C_r(X)$ to the $\tilde C_{D-r}(X)$ of the dual lattice, which we will label $\star$. The $D-r$ cell $\tilde c^{D-r}=\star c^{r}$ is taken to pierce $c^r$ such that the orientation of the direct product of tangent space of $c^r$ and $\tilde c^{D-r}$ matches that of the tangent space $X$ at the point of intersection. We note that $\star^2c^{r}=(-1)^{r(D-r)}c^r$. 

We can now compose the $\star$-operator and $\partial$ to construct the co-boundary operator $\hat\partial$ which maps
\be
\hat\partial: C_r(X)\rightarrow C_{r+1}(X)
\ee
where we define
\be
\hat\partial c^r \equiv (-1)^{(D-r-1)(r+1)}\star\partial\star c^r\;.
\ee
which is equvalent to the statement
\be
\star\hat\partial=\partial\star
\ee
Note that $\partial^2=\hat\partial^2=0$.

An explicit construction of the boundary, co-boundary and $\star$-operators of a cubic lattice is given by
\begin{align}
&\partial c^r_{x;i_1,i_2,\dots, i_r}=\sum_{k}(-1)^{k+1}\left(c^{r-1}_{x+\hat i_k;i_1,i_2,\dots \overset{\circ}{i}_k \dots i_{r}}-c^{r-1}_{x;i_1,i_2,\dots \overset{\circ}{i}_k \dots i_{r}}\right)\;,\\
&\hat\partial c^{r}_{x,i_1,i_2,\dots,i_r}=\sum_{j\ne i_1,i_2,\dots,i_r}(c^{r+1}_{x,i_1,i_2,\dots,i_r,k}-c_{x-\hat j,i_1,i_2,\dots,i_r,k})\;,\\
&\star c^r_{x,i_1,i_2,\dots,i_r}=\frac{1}{(D-r)!}\sum_{i_{r+1}',i_{r+2}',\dots,i_{D}'}\epsilon_{i_1,i_2,\dots, i_{r},i_{r+1}',i_{r+2}',\dots, i_{D}'}\tilde c^{D-r}_{x+\hat s-\hat i_{r+1}'-\dots-\hat r_{D}',i_{r+1}',\dots,i_{D}'}\;,\\
&\star \tilde c^{r}_{\tilde x,i_1,i_2,\dots, i_{r}}=\frac{1}{(D-r)!}\sum_{i_{r+1}',i_{r+2}',\dots,i_{D}'}\epsilon_{i_1,i_2,\dots, i_{r},i_{r+1}',i_{r+2}',\dots, i_{D}'}\tilde c^{D-r}_{\tilde x-\hat s+\hat i_{r+1}'+\dots+\hat r_{D}',i_{r+1}',\dots,i_{D}'}\;,
\end{align} 
where we labeled a cubic $c^{r}$ cell with one of its vertices, and the spatial directions $i_1,i_2,\dots,i_r$, with $i_k=1,2,\dots, r$, and where $\hat i_k$ is a unit lattice vector in the direction $i_k$, $\hat s=\frac{\hat 1+\hat 2+\dots+\hat D}{2}$ is the vector which translates a cubic lattice to its dual-lattice (also cubic), while the $\circ$ indicates that the index is omitted. 

Operators can live on these $r$-cells. Let $A_{c^r}$ be an operator on an $r$-cell $c^r$ which we will call an $r$-form operator (or an $r$-cochain operator). We can then define a map from an $r$-form operator to an $(r+1)$-form operator by an exterior derivative
\be\label{eq:exterior_der}
(dA)_{c^{r+1}}\equiv \sum_{c^r\in\partial c^{r+1}}A_{c^{r}}\;.
\ee
Note that $d^2=0$. Similarly we define a divergence operator, which maps an $r$-form operator to an $(r-1)$-form operator
\be\label{eq:divergence}
(\delta A)_{c^{r-1}}\equiv \sum_{c^{r}\in\hat\partial c^{r-1}}A_{c^{r}}\;.
\ee
We will also define a map $\star$ which maps an operator $A_{c^r}$ on $c^{r}$ into an operator $(\star A)_{\tilde c^{D-r}}$on $\tilde c^{D-r}$ as follows
\be
(\star A)_{\tilde c^{D-r}}\equiv A_{\star \tilde c^{D-r}}\;.
\ee
Let now $A_{c^{r}}$ be an $r$-form on the lattice while $B_{\tilde c^{D-r-1}}$ be an $D-r-1$ form on the dual-lattice. We have that, if $X$ is a closed manifold
\be
\sum_{c^{r+1}} (dA)_{c^{r+1}}B_{\star c^{r+1}}=(-1)^{r+1}\sum_{c^{r}}A_{c^r} (dB)_{\star c^{r}}\;.
\ee
We will also make use of the slightly modified version of the Kronecker delta
\be
\delta_{c^r,c^{'r}}=\begin{cases}{}
1 & \text{if $c^r=c^{'r}$}\\
-1 &\text{if $c^r=-c^{'r}$}\\
0 &\text{otherwise}
\end{cases}
\ee

Let us briefly rewrite the theory \eqref{eq:H1p1d} in this notation. We define operators $\phi_x$ and $n_l$ on the sites $x$ and links $l$ respectively. We write the Hamiltonian
\be
H=\sum_{x} \frac{1}{2aJ}\pi_x^2+\sum_l \frac{J}{2a}((d\phi)_l+2\pi n_l)^2\;.
\ee
We define $\tilde\phi_{\tilde x}$ to be an operator on the dual lattice conjugate to $n_l$, with the commutation relations
\be
[\tilde \phi_{\star l},n_{l'}]=i\delta_{l,l'}\;.
\ee

\subsection{$U(1)$ gauge theory in 2 spatial dimensions}

We now consider a U(1) gauge theory on a spatial lattice. We define such a theory with gauge fields $A_l$ on spatial links $l$ of the 2d lattice $\Lambda$. We construct a Hamiltonian
\be
H=\sum_{l}\frac{1}{2\beta a^2} \pi_l^2+\sum_p\frac{\beta}{2a^2}[(dA)_p+2\pi n_p]^2\;,
\ee
where $\pi_l$ is the canonical momentum conjugate to $A_l$, and where
\be
(dA)_p= \sum_{l\in \partial p}A_l\;,
\ee
is the exterior derivative. We also added an operator on the plaquette $n_p$ with an integral spectrum, which is needed to interpret $A_l$ as a compact gauge field. Indeed we must have that $A_l\rightarrow A_l+2\pi k_l$, for some integers $k_l$ to be a gauge symmetry. In addition, we impose the Gauss law constraint
\be
(\delta \pi)_x= \sum_{l\in\hat\partial l} \pi_l=0\;,
\ee
where $\delta$ is the lattice divergence operator \eqref{eq:divergence}. 

The introduction of $n_p$ integer valued operator implies the existence of a conjugate operator $\phi_{\tilde x}$ which we take to live on the dual lattice site. We impose the commutation relation
\be
[\phi_{\tilde x},n_p]=i\delta_{\star \tilde x,p}=i\delta_{\tilde x,\star p}
\ee

We impose the gauge symmetry
\begin{align}
&A_l\rightarrow A_l+2\pi k_l\;,\\
&n_p\rightarrow n_p-2\pi (dk)_p\;,
\end{align}
which is generated by the operator
\be
U[k]=e^{i\sum_{l}2\pi \pi_l k_l+i\sum_{p}(dk)_p \phi_{\star p}}=e^{i\sum_{l}(2\pi \pi_l+(d\phi)_{\star l})k_l}\;,
\ee
The requirement that every physical state is invariant under $U[k]$ implies that (the minus sign on the r.h.s. is for convenience)
\be
\pi_l+\frac{(d\phi)_{\star l}}{2\pi}=-m_{\star l}
\ee
where $m_{\tilde l}$ is an operator on the dual links with an integral spectrum. Since we assume that $\phi_{\tilde x}$ and $A_{l}$ commute, we must also impose
\be\label{eq:Atn_comm}
[A_{l}, m_{\tilde l}]=i\delta_{l,\star \tilde l}=-i\delta_{\star l,\tilde l}\;,
\ee
i.e. $m_{l}$ now serves as the conjugate momentum of $A_l$. Therefore
\begin{equation}
\sum_l \pi_l^2= \frac{1}{(2\pi)^2}\sum_{\tilde l}[(d\phi)_{\tilde l}+2\pi  m_{\tilde l}]^2
\end{equation}
We finally get that  the Hamiltonian is now
\be
H=\sum_{\tilde l} \frac{1}{\beta(2\pi)^2} \left((d\phi)_{\tilde l}+2\pi  m_{\tilde l}\right)^2+\sum_p \frac{\beta}{2}\left(F_p+2\pi n_p\right)^2\;.
\ee
Note that we had that $n_p$ commutes with $\pi_l$, but since $[\phi_{\tilde x},n_p]=i\delta_{\tilde x,\star p}$ we have that
\be
[n_p,(d\phi)_{\tilde l}+2\pi  m_{\tilde l}]=0\;.
\ee
From the above equation we have that
\be
[n_p, m_{\tilde l(\tilde x,\tilde y)}]=\frac{1}{2\pi}i(\delta_{\tilde x,\star p}-\delta_{\tilde y,\star p})
\ee
where $\tilde l(\tilde x,\tilde y)$ denotes a dual link  which starts at $\tilde x$ and ends at $\tilde y$. Further, the Gauss law constraint translates into
\be
(d m)_{\tilde p}=0\;.
\ee
We could further label $\Pi_{\tilde x}=\frac{F_{\star \tilde x}+2\pi n_{\star \tilde x}}{2\pi}$. Note that $\Pi_{\tilde x}$ serves as the conjugate momentum to $\phi_{\tilde x}$, i.e.
\be
[\phi_{\tilde x},\Pi_{\tilde x'}]=[\phi_{\tilde x},n_{\star\tilde x'}]=i\delta_{\tilde x,\tilde x'}\;.
\ee
Moreover $\Pi_{\tilde x}$ commutes with $ m_{\tilde l}$. To see this, note that
\be\label{eq:inter0}
[\Pi_{\tilde x}, m_{\tilde l(\tilde y,\tilde z)}]=\frac{1}{2\pi}[F_{\star \tilde x}, m_{\tilde l(\tilde y,\tilde z)}]+[n_{\star \tilde x}, m_{\tilde l(\tilde y,\tilde z)}]=\frac{1}{2\pi}[F_{\star \tilde x}, m_{\tilde l(\tilde y,\tilde z)} ]+\frac{i}{2\pi}(\delta_{\tilde y,\tilde x}-\delta_{\tilde z,\tilde x})\;.
\ee
Now we write
\be
[F_{\star\tilde x}, m_{\tilde l(\tilde y,\tilde z)}]=\sum_{l\in\partial \star\tilde x}[A_l, m_{\tilde l(\tilde y,\tilde z)}]=-i\sum_{\tilde l'\in \hat\partial\tilde x}\delta_{\tilde l',\tilde l(\tilde y,\tilde z)}=-i\delta_{\tilde x,\tilde y}+i\delta_{\tilde x,\tilde z}\;.
\ee
In going from the second to the third step above we used the fact that $l\in \partial\star \tilde x$ is the same as $\star l \in -\hat\partial\tilde x$ and, writing $A_l= -A_{\star (\star l)}$ we replaced the sum over $l$ by the sum over $\tilde l=\star l$. So, combining the above with \eqref{eq:inter0} we have that
\be
[\Pi_{\tilde x}, m_{\tilde l}]=0
\ee
The Hamiltonian then becomes
\be\label{eq:H_dual_photon}
H=\sum_{\tilde x} \frac{\beta(2\pi)^2}{2}\Pi_{\tilde x}^2+\sum_{\tilde l} \frac{1}{\beta(2\pi)^2}\left((d\phi)_{\tilde l}+2\pi  m_{\tilde l}\right)^2\;,
\ee
which is the Villain Hamiltonian of the compact scalar on the dual lattice. Note that we have an additional constraint $d\tilde n=0$. This is a no-vortex constraint. 

The no-vortex constraint looks peculiar at first. Surely we could think of the above Hamiltonian without this constraint. This theory has an integer-spectrum operator $ m_{\tilde l}$, living on dual links. As such, its natural conjugate momentum is an angle-valued operator, living on the dual links $\tilde l$ or, equivalently, living on original links $l$, which we label $A_l$. Now the constraint $d m=0$ simply comes from demanding gauge invariance $A_l\rightarrow A_l+(d\lambda)_l$, i.e. it is a Gauss-law constraint.

But what forces us to impose this gauge invariance? We could also consider the Villain Hamiltonian of a compact scalar without such invariance of the link field $A_l$? Notice however that the equations of motion for $m_{\tilde l}$ are
\be
\dot{m}_{\tilde l}=0\;.
\ee 
So the $\tilde n$ operator is in a sense not dynamical, and if we have a state which has a vortex on the dual plaquette $(dm)_{\tilde p}\ne 0$, then that vortex will be there for all other times. Hence the Hilbert space of such a theory decomposes into superselection sectors. One can just as well consider the theory to have a constraint $(dm)_{\tilde p}=0$, and consider the other superselection sectors as temporal (Wilson) line-operator insertions imposing a different superselection sector.

\subsection{$U(1)$ gauge theory in 3 spatial dimensions and electric-magnetic duality} 

Now consider the Hamiltonian
\be
H= \frac{1}{2\beta}\sum_l ({\pi^e}_l)^2+\sum_{p} \frac{\beta}{2}((dA^e)_p+2\pi n_p)^2\;,
\ee
where the spatial lattice is three dimensional. We use the superscript $e$ to label the electric gauge field and its canonical momentum. The operator $n_p$ again has an integral spectrum, and hence we associate a canonical conjugate operator $A^m_{\tilde l}$, living on the dual lattice link as follows
\be
[A^m_{\star p},n_{p'}]=i\delta_{p,p'}\;.
\ee
The operator $A^m_{\tilde l}$ will be interpreted as the dual (magnetic) gauge field. We impose the gauge invariance condition
\be
A^m_{\tilde l}\rightarrow A^m_{\tilde l}+(d\lambda)_{\tilde l}\;,
\ee
where $\lambda_{\tilde x}$ is a gauge parameter on the dual-lattice site. The above transformation is implemented by an operator
\be
e^{i\sum_{\tilde l}n_{\star \tilde l} (d\lambda)_{\tilde l}}=e^{-i\sum_{\tilde x}(dn)_{\star \tilde x}\lambda_{\tilde x}}\;.
\ee
The above operator must be an identity operator on the physical states for any choice of $\lambda_{\tilde x}$, so we must have that $(dn)_{c}=0$ on any cube $c$ of the spatial lattice. This is the no-monopole constraint. Similarly as before, if we wish to consider the temporal monopole line operators, then the constraint should be modified to be different from zero at some cubes $c$ corresponding to the dual lattice sites $\tilde x$ where the static probe monopole lives. 

By the same argument for gauge symmetry of $A^e_l$, we have that $(\delta \pi)_x=0$ -- the Gauss law constraint. Now we must implement the discrete gauge symmetry constraints
\begin{align}
&A^e_{l}\rightarrow A^e_l+2\pi k_l\;,\\
&n_p \rightarrow n_p - (dk)_p\;.
\end{align}
The above is implemented by
\be
e^{i\sum_l 2\pi k_l \pi_l+i\sum_{p}(dk)_pA^m_{\star p}}\;,
\ee
which, again, has to act as identity on the physical states. This implies that
\be
\pi^e_l = \frac{1}{2\pi} \left(-(dA^m)_{\star l}+2\pi m_{\star l}\right)\;,
\ee
where $m_{\tilde p}$  is an operator on the dual plaquette with the integer spectrum. Moreover we must have that
\be
[A^e_l,\pi^e_{l'}]=i\delta_{l,l'}\Rightarrow [A^e_l,m_{\star l'}]=i\delta_{l,l'}\;.
\ee
Similarly like before we note that since we assumed that $\pi_l$ commutes with $n_p$, we must have that
\be
[\pi^e_{\star \tilde l},n_p]=-\frac{1}{2\pi}[(dA^m)_{\tilde p},n_p]+[m_{\tilde p},n_p]=0
\ee
so that
\be
[m_{\tilde p},n_p]=\frac{1}{2\pi}\sum_{\tilde l\in \partial \tilde p}[A^m_{\tilde l},n_p]=\frac{1}{2\pi}\sum_{\tilde l \in \partial \tilde p}i\delta_{\star\tilde l,p}=\frac{iL(\partial p,\partial \tilde p)}{2\pi}\;,
\ee
where $L(\partial p,\partial \tilde p)$ is the linking number between the boundary of the plaquette $p$ and the boundary of the dual plaquette $\tilde p$.
Moreover we define
\be
\pi^m_{\tilde l}=\frac{1}{2\pi}\Big((dA^e)_{\star \tilde l}+2\pi n_{\star\tilde l}\Big)\;.
\ee
The operator above acts like a canonical momentum of $A_{\tilde l}^m$
\be
[A^m_{\tilde l},\pi^m_{\tilde l'}]=i\delta_{\tilde l,\tilde l'}\;.
\ee
Moreover $\pi_{\tilde l}^m$ commutes with $m_{\tilde p}$
\be
[m_{\tilde p},\pi^m_{\tilde l}]=\frac{1}{2\pi}[m_{\tilde p},(dA^e)_{\star\tilde l}]+[m_{\tilde p},n_{\star \tilde l}]\;.
\ee
Indeed since
\be
[m_{\tilde p},(dA^e)_{p}]=-i\sum_{l\in\partial p}\delta_{\star \tilde p,l}=-iL(\partial p,\partial\tilde p)\;,
\ee
hence
\be
[m_{\tilde p},\pi_{\tilde l}^m]=0\;.
\ee
Finally we have the dual form of the Hamiltonian
\be
H=\sum_{\tilde l}\frac{\beta (2\pi)^2}{2}(\pi^m_{\tilde l})^2+\frac{1}{2\beta(2\pi)^2}\sum_{\tilde p}\left((dA^m)_{\tilde p}-2\pi m_{\tilde p}\right)^2\;.
\ee
Now assume that the lattice $\Lambda$ is a hypercubic lattice, and define a translation map $f$ which maps the lattice $\Lambda$ to its dual and  $\tilde\Lambda$ to the $\Lambda$. We can then redefine the operators
\begin{align}
&{A'_l}^e=-A^m_{f(l)} && n_p'=m_{f(p)}\\
&{A'_{\tilde l}}^m=A_{f(\tilde l)}^e && m_{\tilde p}'=-n_{f(\tilde p)}
\end{align}

Note now that models of this sort can be coupled to both magnetic as well as electric matter in a standard way, just like in the space-time counterparts \cite{Sulejmanpasic:2019ytl,Anosova:2022yqx}.

\subsection{$p$-form $U(1)$ gauge theory in $D$ dimensions}

A $p$-form gauge theory consists of $p$-form (or a $p$-cochain) operator $A_{c^p}$ living on a $p$-cell $c^p$. The canonical momentum to $A_{c^p}$ is given by $\Pi_{c^p}$. In $D$ spatial dimensions, we formulate its Hamiltonian as
\be
H=\sum_{c^{p}}\frac{1}{2\beta a}\Pi_{c_p}^2+\sum_{c^{p+1}}\frac{\beta}{2a}((dA)_{c^{p+1}}+2\pi n_{c^{p+1}})^2\;,
\ee
where $n_{c^{p+1}}$ is a $p+1$-form, integer valued operator, whose canonical dual (coordinate) we will take to live on the dual lattice, i.e. $A^m_{\tilde c^{D-p-1}}$, such that
\be\label{eq:Amn_comm}
[A^m_{\star c^{p+1}},n_{c^{'p+1}}]=i (-1)^{(p+1)(D-p-1)}\delta_{c^{p+1},c^{'p+1}}\;.
\ee
or, equivalently
\be
[A^{m}_{\tilde c^{D-p-1}},n_{c^{p+1}}]=i \delta_{c^{p+1},\star \tilde c^{D-p-1}}=i(-1)^{(p+1)(D-p-1)}\delta_{\star c^{p+1},\tilde c^{D-p-1},}
\ee
As before, the Kronecker delta is defined such that it is $+1$ if the two cells are the same with the same orientation, $-1$ if they are the same with opposite orientation and $0$ if they are distinct. 
The Hamiltonian is invariant under a gauge transformation 
\be
A_{c^p}\rightarrow A_{c^p}+(d\lambda)_{c^p}\;.
\ee
which, when we impose the neutrality of the physical states under the transformation, leads to a Gauss constraint 
\be
(\delta \Pi)_{c^{p-1}}=0\;.
\ee
Similarly we can impose the gauge symmetry
\begin{align}
&A^e_{c^p}\rightarrow A_{c^p}+2\pi k_{c^p}\;,\\
&n_{c^{p+1}}\rightarrow n_{c^{p+1}}-(dk)_{c^{p+1}}\;,
\end{align}
which is implemented by an operator
\be
e^{i2\pi \sum_{c^p}k_{c^p}\Pi^e_{c^p}+(-1)^{(p+1)(D-p-1)}\sum_{c^{p+1}}(dk)_{c^{p+1}}A^m_{\star c^{p+1}}}=e^{i\sum_{c^p}k_{c^p}(2\pi\Pi^e_{c^p}+(-1)^{pD-D}(dA^m)_{\star c^{p}})}
\ee
which leads to the constraint
\be
\Pi_{c^{p}}=-\frac{1}{2\pi }(-1)^{(p-1)D}(d A^m)_{\star c^{p}}+m_{\star c^p}\;,
\ee
where $m_{\tilde c^{D-p}}$ is an integer spectrum operator, living on the $D-p$ cells of the dual lattice. Note that this means that
\be
[A^e_{c^p},m_{\star  c^{'p}}]=i\delta_{c^p,c^{'p}}\Leftrightarrow [A_{\star \tilde c^{D-p}}^e,m_{\tilde c^{'D-p}}]=(-1)^{p(D-p)}i\delta_{\tilde c^{D-p},\tilde c^{'D-p}}\;,
\ee
which is the mirror image of \eqref{eq:Amn_comm} and can also be written as
\be
[A^e_{c^p},m_{\tilde c^{D-p}}]=i\delta_{\star c^p,\tilde c^{D-p}}=i(-1)^{p(D-p)}\delta_{c^p,\star \tilde c^{D-p}}\;.
\ee
Recall that we take $\Pi_{c^p}$ to commute with the field $n_{c^{p+1}}$, and hence
\be
-(-1)^{(p-1)D}[(dA^m)_{\tilde c^{D-p}},n_{c^{p+1}}]+2\pi[m_{\tilde c^{D-p}},n_{c^{p+1}}]=0
\ee
or\footnote{We used $\delta_{\star \tilde c^{D-p-1},c^{p+1}}=\delta_{\star^2\tilde c^{D-p-1},\star c^{p+1}}=(-1)^{(D-p-1)(p+1)}\delta_{\tilde c^{D-p-1},\star c^{p+1}}$}
\begin{multline}
[m_{\tilde c^{D-p}},n_{c^{p+1}}]=\frac{(-1)^{pD-D}}{2\pi}\sum_{\tilde c^{D-p-1}\in \partial \tilde c^{D-p}}\underbrace{[A_{\tilde c^{D-p-1}}^m,n_{c^{p+1}}]}_{=i(-1)^{(p+1)(D-p-1)}\delta_{\tilde c^{D-p-1},\star c^{p+1}}}=\\=\frac{(-1)^{p+1}}{2\pi}\sum_{\tilde c^{D-p-1}\in \partial\tilde c^{D-p}}\delta_{\tilde c^{D-p-1},\star c^{p+1}}\;.
\end{multline}

Now recall that $\star c^{p+1}$ is a $D-p-1$-cell which pierces $c^{p+1}$ cell in such a way that the induced orientation on the $D$-cell which is obtained by the extension of $c^{p+1}$ by $\tilde c^{D-p-1}$ is standard\footnote{By a standard orientation we mean the orientation given by the ordering of the lattice coordinates $1,2,3,\dots D$.}. In other words the Kronecker delta picks up a positive contribution whenever $D-p-1$-cell $\tilde c^{D-p-1}$ pierces the $c^{p+1}$ cell, such that $c^{p+1}$ with $\tilde c^{D-p-1}$ form a standard orientation, and negative if the piercing is opposite. We can hence define the linking number between the boundary two cells as
\be
L(\partial c^{p+1},\partial \tilde c^{D-p})=\sum_{\tilde c^{D-p-1}\in \partial \tilde c^{D-p}}\delta_{ \tilde c^{D-p-1},\star c^{p+1}}\;.
\ee
This means that
\be
[n_{c^{p+1}},m_{\tilde c^{D-p}}]=\frac{(-1)^{p+1}}{2\pi}L(\partial c^{p+1},\partial\tilde c^{D-p})\;.
\ee
Hence we can write the Hamiltonian as
\be
H=\frac{1}{2\beta(2\pi)^2}\sum_{\tilde c^{D-p}}((dA^m)_{\tilde c^{D-p}}-(-1)^{pD-D}2\pi m_{\tilde c^{D-p}})^2+\frac{\beta}{2}\sum_{c^{p}}((dA^e)_{c^{p+1}}+2\pi n_{c^{p+1}})^2\;.
\ee
Further we can also define the dual momentum $\Pi^m_{\tilde c^{D-p-1}}$ of $A^m_{c^{D-p-1}}$ as
\be
\Pi^m_{\tilde c^{D-p-1}}=(dA^e)_{\star \tilde c^{D-p-1}}+2\pi n_{\star \tilde c^{D-p-1}}\;.
\ee
We can check the commutation relations of $\Pi^m_{\tilde c^{D-p-1}}$ with $m_{\tilde c^{D-p}}$. We have that
\begin{multline}
[\Pi_{\star c^{p+1}}^m,m_{\tilde c^{D-p-1}}](-1)^{(p-1)(D-p-1)}=\frac{1}{2\pi}[dA^{e}_{c^{p+1}},m_{\tilde c^{D-p}}]+ [n_{c^{p+1}},m_{\tilde c^{D-p}}]=\\
=\frac{1}{2\pi}[dA^{e}_{c^{p+1}},m_{\tilde c^{D-p}}]-i(-1)^pL(\partial c^{p+1},\partial\tilde c^{D-p})\\;.
\end{multline}
Since
\begin{multline}
[(dA^e)_{c^{p+1}},m_{\tilde c^{D-p}}]=\sum_{c^p\in \partial c^{p+1}}\underbrace{[A^e_{c^{p}},m_{\tilde c^{D-p}}]}_{ i(-1)^{p(D-p)}\delta_{c^p,\star \tilde c^{D-p}}}=\\=i(-1)^{p(D-p)}L(\partial \tilde c^{D-p},\partial c^{p+1})=i(-1)^{p}L(\partial c^{p+1},\partial \tilde c^{D-p})
\end{multline}
where we used that $L(\partial c^{p+1},\partial \tilde c^{D-p})=(-1)^{pD}L(\partial \tilde c^{D-p},\partial c^{p+1})$ shown in the Appendix \ref{app:linking}.

\subsection{Comments on the BF theories}\label{sec:BF_theory}

We finally give brief comments on the BF theories. Such theories have a zero Hamiltonian, but a nontrivial algebra. We consider a general case of a $p$-form operator $A_{c^p}$ and its counterpart $B_{\tilde c^{D-p}}$. We impose the following commutation relations 
\be
\left[{A_{c^p}},{B_{\tilde c^{D-p}}}\right]=\frac{i2\pi \delta_{c^p,\star\tilde c^{D-p}}}{N}\;,
\ee
where $N$ is a positive integer. We further impose a gauge symmetry $A_{c^p}\rightarrow A_{c^p}+(d\lambda)_{c^p}$ with $\lambda-{c^{p-1}}$ a real, $(p-1)$-form  gauge parameter. This symmetry is implemented by an operator
\be
e^{\frac{iN}{2\pi}\sum_{c^{p}}(d\lambda)_{c^p}{B_{\star c^p}}}=\mathbb I\;,
\ee
which results, upon partial integration, in the constraint $(dB)_{\tilde c^{D-p+1}}=0$, i.e. $B$ is a flat operator. Similarily by imposing the gauge symmetry of $B\rightarrow B+d\lambda$, we get that $(dA)_{c^{p+1}}=0$. Further, we also want to impose that $A_{c^p}\rightarrow A_{c^p}+2\pi k_{c^{p}}$, and $B_{\tilde c^{D-p}}\rightarrow B_{\tilde c^{D-p}}+2\pi k_{\tilde c^{D-p}}$ with $k_{c^{p}},k_{\tilde c^{D-p}}\in\mathbb Z$ we get that
\be
e^{i N A_{c^{p}}}=e^{i N B_{\tilde c^{D-p}}}=\mathbb I\;,
\ee
which indicates that $A_{c^{p}}$ and $B_{\tilde c^{D-p}}$ are $\mathbb Z_{N}$ gauge fields. Moreover note that the constraint on $dB$ and $dA$ should now be interpreted as a mod $2\pi$ constraint, i.e. as 
\be
e^{i (dB)_{\tilde c^{D-p+1}}}=e^{i(dA)_{c^{r+1}}}=\mathbb I\;.
\ee

It is now easy to see that Wilson sheets of $A$ and $B$ have anyonic statistics
\be
e^{iq_1 \sum_{c^p\in C_1}A_{c^p}}e^{iq_2\sum_{\tilde c^{D-p}\in C_2}B_{\tilde c^{D-p}}}=e^{-q_1q_2\sum_{c^p,\tilde c^{D-p}}[A_{c^p},B_{\tilde c^{D-p}}]} e^{-i\frac{2\pi}{N}I(C_2,C_1)}
\ee
where $I(C_2,C_1)$ is the intersection number of the hyper surface $C^1$ with $C_2$ defined as 
\be
I(C_2,C_1)=\sum_{c^p\in C^1,\tilde c^{D-p}\in C^2}\delta_{c^p,\star \tilde c^{D-p}}\;.
\ee
A surface operator in space time $e^{i\oint A}$ which winds in the temporal direction must modify the Gauss constraint as follows. Firstly, note that the component of $A$ which points in time would naturally be intepreted as an object living on the $c^{p-1}$ of the lattice. The operator $e^{i\oint A}$ which spans in time for a fixed spatial $p-1$ hyper-surface $S$ can be seen as modifying the Hilbert space as follows
\be
e^{i(dB)_{\star c^{r-1}}}=e^{-i\frac{2\pi q_1}{N}\sum_{c^{'r-1}\in S}\delta_{c^{r-1},c^{'r-1}}}\;.
\ee
This will guarantee the topological correlation functions between ``loops'' of $B$ and the surface $S$. 

\subsection{Coupling to gauge fields, anomalies and the Ising duality}

Let us now discuss 1+1d theories with scalars coupled to gauge fields. This is well known to be solvable in continuum and we will see that we can construct lattice models which are also solvable. 
Moreover we will explore the 't Hooft anomaly which arises, and discuss why gauging some of the symmetries may be inconsistent.

Let us start with the simplest model: the compact boson Hamiltonian \eqref{eq:H1p1d}. We introduce the gauge fields $A_l$ on links gauging the $\phi_x\rightarrow \phi_x+\alpha$ shift symmetry
\be\label{eq:H_cb_gf}
H=\frac{1}{2J}\sum_x\pi_x^2+\frac{J}{2}\sum_l(d\phi+2\pi n+qA)_l^2+\sum_l \frac{e^2}{2}\left(\Pi_l+\frac{\theta}{2\pi}\right)^2\;,
\ee
where we have decided to gauge the symmetry with a charge $q$, and where $\Pi_l$ is the conjugate momentum to $A_l$. We also introduced the $\theta$-angle.
What about the shift symmetry $\tilde\phi_{\tilde x}\rightarrow \tilde \phi_{\tilde x}+\tilde \alpha$? Since we have the commutation relation
\be
[\tilde \phi_{\tilde x},n_{\star \tilde y}]=i\delta_{\tilde x,\tilde y}\;,
\ee
naively the conserved charge that implements the shift symmetry of $\tilde\phi_{\tilde x}$ is just given by $\sum_{l}n_l$. This however is not gauge invariant under the new discrete symmetry\footnote{We could just not impose this symmetry, but then $A_l$ would be and $\mathbb R$ gauge field, not a $U(1)$ gauge field.} $A_{l}\rightarrow A_l+2\pi k_l$ and $n_l\rightarrow n_l-qk_l$. So the conserved charge should be 
\be
\tilde Q=\sum_l\left(n_l+q\frac{A_l}{2\pi}\right)\;.
\ee
The above charge, however, is no longer conserved. Indeed we have that
\be\label{eq:dotQ}
\dot {\tilde Q}=i[H,\tilde Q]=qe^2\sum_l\Pi_l\;.
\ee
The equations of motion for $A_l$ however also give that
\be
\dot A_l=i[H,A_l]=qe^2\Pi_l\;,
\ee
so we can write
\be
\dot{\tilde Q}=q\partial_t\sum_l A_l\;,
\ee
This is the famous mixed anomaly between the momentum and winding symmetries. Before continuing to solve this gauged model\footnote{This is a bosonized version of the charge $q$ Schwinger model which has been of interest in some recent literature \cite{Anber:2018jdf,Misumi:2019dwq,Cherman:2022ecu}.}, let us consider gauging only the $\mathbb Z_N$ subgroup of the $U(1)$ symmetry. To do this we let the Hamiltonian be
\be\label{eq:H_cb_gf_ZN}
H=\frac{1}{2J}\sum_x\pi_x^2+\frac{J}{2}\sum_l(d\phi+2\pi n+A)_l^2\;,
\ee
where now we have that $A_l$ is the $\mathbb Z_N$ gauge field discussed in Sec.~\ref{sec:BF_theory}. The conserved dual charge is given by
\be
\tilde Q=\sum_{l}\left(n_l+\frac{A_l}{2\pi}\right)
\ee
which is now still conserved. But notice that it is not an integer, so there is still an anomaly between a discrete $\mathbb Z_N$ momentum symmetry and the $U(1)$ winding symmetry. Now let us try to preserve only a subgroup $\mathbb Z_M$ of the $U(1)$ winding symmetry. Before gauging the $\mathbb Z_{N}$ momentum symmetry, the generator of the $\mathbb Z_{M}$ winding symmetry was
\be
G^M=e^{i\frac{2\pi}{M}\sum_{l}n_l}\;.
\ee
Upon gauging the $\mathbb Z_{N}$ momentum symmetry the above is not gauge invariant under $A_l\rightarrow A_l-2\pi k_l$ and $n_l\rightarrow n_l+k_l$. We want to attach an improperly quantized Wilson line $e^{i\frac{pN}{M}\sum_l A_l}$ with $p\in \mathbb Z$ so that we preserve the property $G^M=\mathbb I$. So let's define
\be
G= e^{\frac{i}{M}\sum_{l}(2\pi n_l+pN A_l)}\;.
\ee
Now the above combination must be gauge invariant under $n_l\rightarrow n_l+k_l$ and $A_l-2\pi k_l$ which can only be true if $pN=1\bmod M$. This condition can only be solved for $p$ if $\GCD(N,M)=1$. This is indeed what one expects in the continuum\footnote{In the continuum, one can put the background $\mathbb Z_N$  gauge fields $\tilde A$  for the $\mathbb Z_{N}$ subgroup of the winding symmetry by the minimal coupling term $\frac{1}{2\pi}\int\tilde A\wedge d\phi$ in the action. Now upon putting background $\mathbb Z_M$ gauge field for the $\phi$ shift symmetry, the minimal coupling term becomes $\frac{1}{2\pi}\int\tilde A\wedge (d\phi+A)$. This renders the term no longer gauge invariant under the large gauge transformations of $\tilde A$ because $\int(d\phi+A)$ is quantized in units of $2\pi/M$. One can however introduce a counter-term $pN\int A\wedge \tilde A$, which does not spoil the gauge invariance of $A$, and $p$ can be picked so that it fixes the gauge non-invariance of $\tilde A$ if $\GCD(M,N)=1$. See \cite{Komargodski:2017dmc,Komargodski:2017smk,Kikuchi:2017pcp} for related discussions.}. 

The story can be repeated for $p$-form gauge fields in arbitrary dimensions, where the two $U(1)$ symmetries are $p$-form and the $(D-p-1)$-form, with a mixed 't Hooft anomaly between them. Again one can show that two discrete subgroup $\mathbb Z_{N}$ and $\mathbb Z_{M}$ do not have a mixed anomalies only if $\GCD(N,M)=1$.

Now let's go back the discussion of the theory \eqref{eq:H_cb_gf}. Notice that the transformation 
\begin{align}
&n_l\rightarrow n_l+qk_l
&A_l\rightarrow A_l-2\pi k_l
\end{align}
is a gauge symmetry, and hence the operator which implements it must be an identity operator
\be
e^{i \sum_l k_l(2\pi \Pi_l+q\tilde\phi_{\star l})}=\mathbb I
\ee
so that
\be
\Pi_l= {M_{\star l}}-\frac{q\tilde \phi_{\star l}}{2\pi}\;,
\ee
where $M_{\star l}$ is an integer valued operator, which must have the commutation relation
\be
[A_l,M_{\star l'}]=i\delta_{l,l'}\;.
\ee
In addition the usual Gauss law says that
\be
(\delta\Pi)_x = -q\pi_x\;,
\ee
which translates into 
\be
\pi_x= \frac{(d\tilde\phi)_{\star x}}{2\pi}-\frac{(dM)_{\star x}}{q}\;.
\ee
On the other hand we know that
\be
\pi_x= \frac{(d\tilde\phi)_{\star x}}{2\pi}+\tilde n_{\star x}\;,
\ee
so that
\be
(dM)_{\tilde l}=-q\tilde n_{\tilde l}\;.
\ee
Note that the above equation says that $M$ is constant in space mod $q$. Meaning that $e^{i2\pi M_{\tilde x}/q}$ does not depend on $\tilde x$. As we will see $M$ will label $q$ degenerate vacua. 
Finally we define a gauge invariant canonical momentum $\tilde p_{\tilde x}$ to $\tilde\phi_{\tilde x}$ as
\be
\tilde p_{\tilde x}=\tilde\pi_{\tilde x}+q A_{\star\tilde x}\;,
\ee
which obeys the following non-zero commutation relations
\begin{align}
&[\tilde \phi_{\tilde x},\tilde p_{\tilde y}]=i\delta_{\tilde x,\tilde y}\;,\\
&[\tilde p_{\tilde x},M_{\tilde y}]=q i\delta_{\tilde x,\tilde y}
\end{align}
so the Hamiltonian can then be written as
\be
H=\frac{J}{2}\sum_{\tilde x}\tilde p_{\tilde x}^2+\frac{1}{2J(2\pi)^2}\sum_{\tilde{l}}\left((d\tilde\phi)_{\tilde l}+2\pi \tilde n_{\tilde l}\right)^2+\sum_l \frac{e^2q^2}{2(2\pi)^2}\left(\tilde\phi_{\tilde x}-\frac{2\pi M_{\tilde x}}{q}+\frac{\theta}{q}\right)^2\;.
\ee
Firstly note that the $\theta$ term can just be absorbed into the anomalous shift of $\tilde\phi_{\tilde x}$ as expected. Further, $M_{\tilde x}$ commutes with the Hamiltonian, and can hence be set to a numerical value. The same is true for $\tilde n_{\tilde l}$. We must further impose the constraint that $\tilde n_{\tilde l}=-\frac{(dM)_{\tilde l}}{q}$. But if $\tilde n_{\tilde l}$ is a total derivative, we can absorb it in the shift of $\tilde\phi_{\tilde x}$. The remaining model is then a gapped lattice scalar with mass $\frac{eq\sqrt{J}}{2\sqrt{2}\pi}$. Notice however that the model has $q$ vacua which are distinguished by the operator $e^{i \frac{2\pi N}{q}}=e^{i\frac{2\pi M_{\tilde x}}{q}}$, which is space-independent and defines an integer $M$, well defined mod $q$, which labels the vacua. The $q$ vacua correspond to the degenerate universes associated with the $\mathbb Z_{N}$ $1$-form symmetry. 
Now let us consider the model with dynamical vortices, which are described by operators $e^{\pm  i\tilde \phi_{\tilde x}}$. In particular we have a Hamiltonian
\begin{multline}\label{eq:H_cb_gf_vortex}
H=\frac{J}{2}\sum_{\tilde x}\tilde p_{\tilde x}^2+\frac{1}{2J(2\pi)^2}\sum_{\tilde{l}}\left((d\tilde\phi)_{\tilde l}+2\pi \tilde n_{\tilde l}\right)^2\\+\sum_{\tilde x}\left[ \frac{e^2q^2}{2(2\pi)^2}\left(\tilde\phi_{\tilde x}-\frac{2\pi M_{\tilde x}-\theta}{q}\right)^2+m\cos(\tilde\phi_{\tilde x})\right]\;.
\end{multline}
Diagonalizing $M_{\tilde x}$ we have that the Hamitlonian splits into $q$ sectors labeled by the integer $M=0,1,\dots,q-1$
\be
H_M=\frac{J}{2}\sum_{\tilde x}\tilde p_{\tilde x}^2+\frac{1}{2J(2\pi)^2}\sum_{\tilde{l}}(d\tilde F)_{\tilde l}^2+\sum_{\tilde x} \left[\frac{e^2q^2}{2(2\pi)^2}\tilde F_{\tilde x}^2+m\cos\left(\tilde F_{\tilde x}+\frac{2\pi M-\theta}{q}\right)\right]\;,
\ee
where $\tilde F_{\tilde x}$ is related to $\tilde \phi_{\tilde x}$ as
\be
\tilde F_{\tilde x}=\tilde\phi_{\tilde x}-\frac{2\pi M_{\tilde x}-\theta}{q}\;.
\ee
Now notice that for generic values of $\theta$ and $q$, all vacua labeled by $M$ have a distinct Hamiltonian, and hence a different ground state. When $\theta=\pi$ however, notice that charge conjugation symmetry $C$ which takes $\tilde F_{\tilde x}\rightarrow -\tilde F_{\tilde x}$ acts on $M$ as
\be
M\rightarrow -M+1 \bmod q
\ee
Now if the above symmetry is leaving the vacuum labeled by $M$ invariant, we would have
\be
2M-1=0 \bmod q\;,
\ee
which is only possible if $q$ is odd. Hence for even $q$, all vacua transform under the $C$ symmetry, and, in particular, the ground state must be degenerate. This is the reflection of the mixed anomaly between the $C$-symmetry and the $\mathbb Z_{q}$ 1-form symmetry at $\theta=\pi$ \cite{Komargodski:2017dmc}. 

When $q=1$, there will be an Ising transition at $\theta=\pi$ as $m$ is dialed. If $m$ is large and positive, the Hilbert space is projected onto the states with $\tilde F_{\tilde x}=0$, which does not break the $C$-symmetry. When $m$ is large and negative, $\tilde F_{\tilde x}$ is forced to be either $+\pi$ or $-\pi$, and the $C$-symmetry is broken\footnote{Notice that unlike $\tilde\phi_{\tilde x}$, $\tilde F_{\tilde x}$ is not a compact operator, and $\tilde F_{\tilde x}=\pi$ and $\tilde F_{\tilde x}=-\pi$ are distinct values of the field. }. 

We can also construct another model in the same universality class as the one above. Namely let us consider the following analogous model to \eqref{eq:H_cb_gf}
\be\label{eq:H_cb_gf1}
H=\frac{1}{2J}\sum_x\pi_x^2-J\sum_l\cos((d\phi)_l+qA_l)+\sum_l \frac{e^2}{2}\left(\Pi_l+\frac{\theta}{2\pi}\right)^2\;.
\ee
The model above differs from \eqref{eq:H_cb_gf} in that the Villain form was replaced by the more conventional XY-model/Wilson type. Because of this, the model will not have the winding symmetry, and is hence in the same universality class as \eqref{eq:H_cb_gf_vortex}. We want to study this model in the limit of strong gauge coupling at $\theta=\pi$. In that case we have that the last term enforces a constraint that $\Pi_{l}$ can take only two values $\Pi_{l}=0,1$. We therefore label $\Pi_{l}\rightarrow \frac{1-\sigma_{\star l}^3}{2}$, where $\sigma^3_{\tilde x}$ is the 3rd sigma matrix living on the dual sites $\tilde x$. Since the Gauss law states that $(\delta \Pi)_x=-q\pi_x^2$, we can replace $\pi_x^2\rightarrow \frac{1}{q^24}(d\sigma^3)_{\star x}^2$. Since $(d\sigma^3)_{\tilde l(\tilde x,\tilde y)}=\sigma^3_{\tilde y}-\sigma^{3}_{\tilde x}$, where $\tilde l(\tilde x,\tilde y)$ is the link starting at dual site $\tilde x$ and ending at the dual site $\tilde y$, we have that 
\be
\sum_{x}\pi_x^2\rightarrow -\frac{1}{2q^2}\sum_{\tilde x} \sigma^3_{\tilde x+1}\sigma^3_{\tilde x}+\text{constant terms}\;,
\ee
which is just the Ising coupling. 

Finally the term $\cos((d\phi)_l+qA_{l})$ always takes the state with $\Pi_{l}=0,1$ into a state with different $\Pi_{l}$, if $q>1$, and acts as a zero operator on the projected Hilbert space. $\Pi_{l}=0,1$. Hence we have that the Hamiltonian exactly becomes that of the Ising model
\be
H_{e^2\rightarrow \infty,q>1}= -\frac{1}{4Jq^2}\sum_{\tilde x}\sigma^3_{\tilde x+1}\sigma^3_{\tilde x}
\ee 
which of course has two ground states. If however $q=1$, then the cosine term does not act as a zero operator. Instead it acts as a $\sigma^1_{\tilde x}$ operator, and the resulting Hamiltonian is
\be
H_{e^2\rightarrow \infty,q=1}= -\frac{1}{4J}\sum_{\tilde x}\sigma^3_{\tilde x+1}\sigma^3_{\tilde x}-\frac{J}{2}\sum_{\tilde x}\sigma^1_{\tilde x}\;.
\ee 
This is known as the transverse field Ising model, and it is exactly solvable, with a transition occurring when the ratio of the coefficients of the second term and the first term is equal to $1$, i.e. at $J=1/\sqrt{2}$. If $J<1/\sqrt{2}$, there are two vacua related by the spin flip symmetry (i.e. $C$ symmetry). If $J>1/\sqrt{2}$ the ground state is unique. This is what we expected from the analysis of \eqref{eq:H_cb_gf_vortex} with $q=1$.

Finally we comment that the quantum Ising model can also be obtained in arbitrary dimensions from the generalization of the above story to $D$ spatial dimensions. To that end, let us consider $D-1$-form gauge field $A_{c^{D-1}}$ and couple it to a $D$-form gauge filed $B_{c^{D}}$ as follows
\be\label{eq:H_D-form_cb_gf}
H=\frac{1}{2J}\sum_{c^{D-1}}\pi_{c^{D-1}}^2-\frac{J}{2}\sum_l\cos(dA+qB)_{c^{D}}^2+\sum_{c^{D}} \frac{e^2}{2}(\Pi_{c^{D}}+\frac{\theta}{2\pi})^2\;,
\ee
where $\pi_{c^{D-1}}$ is a conjugate momentum to $A_{c^{D-1}}$, $\Pi_{c^{D}}$ is the conjugate momentum to $B_{c^{D}}$. When $\theta=\pi$ we again, by very similar reasoning, get the Ising model in the limit $e^2\rightarrow\infty$. The Ising spins $\sigma^3_{\tilde x}$ lives on the dual lattice sites. The Hamiltonian is given by
\be
H_{e^2\rightarrow \infty}=-\frac{1}{2Jq^2}\sum_{<\tilde x,\tilde y>}\sigma^3_{\tilde x}\sigma^3_{\tilde y}-\delta_{q,0}\frac{J}{2}\sum_{\tilde x}\sigma^1_{\tilde x}
\ee
This is the Hamiltonian version of the strong-coupling duality \cite{Sulejmanpasic:2020ubo}.

\section{Exotic theories}

In this section we study some exotic fracton models which have subsystem symmetries. In particular we will consider a version of the XY-plaquette model \cite{paramekanti2002ring}. Much like the XY model is an analogue of a compact scalar model, the XY-plaquette model can be seen as an analogue of a model described in the continuum by a Minkowski Lagrangian\footnote{The ``continuum'' theory here is subtle because of the UV/IR mixing, which was the main focus of the works of Seiberg and Shao \cite{Seiberg:2020bhn,Seiberg:2020wsg,Seiberg:2020cxy,Gorantla:2020xap}. }
\be
L= \frac{\mu_0}{2}(\dot\phi)^2-\frac{1}{\mu_1}(\partial_1\partial_2\phi)^2\;.
\ee
This model has a subsystem symmetry associated with the shift $\phi(x^1,x^2)\rightarrow \phi(x^1,x^2)+f(x^1)+g(x^2)$ where $f$ and $g$ can be arbitrary functions of $x^1$ and $x^2$ respectively. This we will call the momentum subsystems symmetry, in analogy to the compact scalar symmetry. The model has also a winding subsystem symmetry associated with the conserved dipole charges\footnote{The charges can be nontrivial because $\partial_{1}\phi$ and $\partial_2\phi$ are only well defined mod $ 2\pi$. These subtleties of the continuum theory have been the central theme of the works of Seiber and Shao \cite{Seiberg:2020bhn,Seiberg:2020wsg,Seiberg:2020cxy}.} $Q_1(x^1)=\frac{1}{2\pi}\int dx^1 (\partial_1\partial_2\phi)$ and $Q_2(x^2)=\frac{1}{2\pi}\int dx^1 (\partial_1\partial_2\phi)$ 
\cite{ma2018higher,Seiberg:2020bhn,Seiberg:2020wsg,Seiberg:2020cxy,Gorantla:2020xap,Gorantla:2020jpy,Gorantla:2021svj,Gorantla:2021bda,Gorantla:2022eem,Gorantla:2022ssr,Burnell:2021reh,distler2022spontaneously}. The winding symmetry can only be emergent in the XY-plaquette model, just like the winding symmetry of the XY-model in (1+1)d only emerges in a particular regime. In \cite{Gorantla:2021svj} a space-time lattice model was constructed which has an exact winding dipole symmetry. Models discussed here are the Hamiltonian analogues of these. 

\subsection{XY-plaquette model with exact winding symmetries}

Consider now the Hamiltonian 
\be\label{eq:H1fracton}
H=\sum_x\left(\frac{1}{2Ja}\pi_x^2+\frac{J}{2a} \left(\Delta_1\Delta_2\phi_x+2\pi n_x\right)^2\right)\;.
\ee
where $x$ is a position vector on the 2d lattice, and $\Delta_{i}\phi_x=\phi_{x+\hat i}-\phi_x$, with $\hat i$ being a unit lattice vector in the spatial direction $i=1,2$. Note $a$ has dimensions of length and $J$ is dimensionless. The operator $n_x$ has an integer spectrum, with a canonical  conjugate $\varphi_x$
\be
[\varphi_x,n_y]=i\delta_{x,y}\;.
\ee
Now we note that the transformation
\begin{align}
&\phi_x\rightarrow \phi_x+2\pi k_x\;,\\
&n_x \rightarrow n_x-\Delta_1\Delta_2 k_x\;,
\end{align}
with $k$ an integer, is an invariance. We want to make the above into a gauge symmetry. The above transformation is generated by an operator
\be
e^{i\sum_x 2\pi k_x \pi_x-i\sum_x(\Delta_1\Delta_2 k)_x \varphi_x}\;.
\ee
we use the ``partial integration'' formula
\be
\sum_x (\Delta_i f)_xg_x=\sum_{x}(f_{x+\hat i}-f_x)g_x= \sum_{x}f_x(g_{x-\hat i}-g_x)=-\sum_x f_x(\Delta_{i}g)_{x-\hat i}\;,
\ee
so we rewrite the generator as
\be
e^{i\sum_x 2\pi k_x \pi_x-i\sum_xk_x \Delta_{1}\Delta_{2}\varphi_{x-\hat 1-\hat 2}}\;.
\ee
The above must be an identity operator on the Hilbert space, so we impose a constraint
\be
\pi_x= \frac{\Delta_1\Delta_2\varphi_{x-\hat 1-\hat 2}}{2\pi}+m_x\;,
\ee
where $m_x$ has an integral spectrum. Moreover since $[n_x,\pi_y]=0$ we have that
\be
[n_x,m_y]=\frac{i}{2\pi}\left(\delta_{x,y}-\delta_{x,y-\hat 2}-\delta_{x,y-\hat 1}+\delta_{x,y-\hat1-\hat2}\right)\;.
\ee
The Hamiltonian takes the form
\be
H=\sum_x\Bigg\{\frac{1}{2Ja(2\pi)^2}\left(\Delta_{1}\Delta_2\varphi_x+2\pi m_{x+\hat 1+\hat 2}\right)^2+\frac{J}{2a}\left(\Delta_1\Delta_2\phi_x+2\pi n_x\right)^2\Bigg\}\;.
\ee
The Hamitlonian is invariant under the replacement
\begin{align}
&\phi_x\rightarrow \varphi_x\;,&& n_x\rightarrow m_{x+\hat 1+\hat 2}\;,\\
&\varphi_x\rightarrow \phi_{x+\hat 1+\hat 2}\;, && m_x\rightarrow n_x\;.
\end{align}
along with $J\rightarrow \left(\frac{1}{2\pi}\right)^2\frac{1}{J}$. This is the self-duality of the model.
The reader can check that the commutation relations
\begin{align}
&[\phi_x,m_y]=i\delta_{x,y}\;, && [\varphi_x,n_y]=i\delta_{x,y}\;,\\
&[n_x,m_y]=\frac{i}{2\pi}\left(\delta_{x,y}-\delta_{x,y-\hat 1}-\delta_{x,y-\hat 2}+\delta_{x,y-\hat1-\hat2}\right)\;,
\end{align}
are invariant under self-dual transformation. Note that, as in the 1+1d counterpart, the square of the self-dual transformation is not identity, but a diagonal lattice translation. The model clearly enjoys two winding symmetries, as the shifts $\phi_x\rightarrow \phi_x+f_2(x^1)+f_2(x^2) $ and $\varphi_{\tilde x}+g_1(x^1)+g_2(x^2)$ where $f_{1,2}$ and $g_{1,2}$ are arbitrary functions of $x^{1,2}$ respectively. The model is also exactly solvable, as we show in the Appendix~\ref{app:XY_sol} and matches nicely the continuum discussion of  \cite{Seiberg:2020bhn}.

\subsection{2+1d Tensor model and the quantum Ising model duality}

Now we want to consider gauging the tensor symmetry which is specified by the current $J_{0,x},J^{12}_x$. We introduce the tensor gauge field $A_{x,0}$ and $A_{x,12}$ with a gauge symmetry 
\begin{align}
&A_{x,0}\rightarrow A_{x,0}+\partial_0\phi_x\;,\\
&A_{x,12}\rightarrow A_{x,12}+\Delta_1\Delta_2\phi_x\;.
\end{align}
We want to construct a theory in which we can identify $A_{x,12}\sim A_{x,12}+2\pi$. Let us consider the Gauge invariant field strength 
\be
F_{x,0,12}=\partial_0A_{12}-\Delta_1\Delta_2A_0\;.
\ee
The (real-time) Lagrangian is given by
\be
L=\sum_x \frac{\beta a}{2}F_{x,0,12}^2\;.
\ee
The Hamiltonian is
\be
H= \sum_x\left((\frac{1}{2a\beta}\Pi_{x,12}^2+(\Delta_1 \Delta_2A_{x,0}) \Pi_{x,12})\right)\;.
\ee
where  $\pi_{x,12}$ as a conjugate momentum to $A_{x,12}$. The conjugate momentum $\pi_{x,0}$ of $A_{x,0}$ is zero (primary constraint in the Dirac constraint classification \cite{dirac2001lectures}), so $\pi_{x,0}$ must commute with the Hamitlonian. This condition gives us, upon ``partial integration'' the secondary constraint, or Gauss law
\be\label{eq:Tensor_gauss}
\Delta_1\Delta_2 \Pi_{x,12}=0\;.
\ee
Since $\pi_{x,0}$ and $\pi_{x,12}$ have a zero Poisson  bracket, the constraints are first class. This is exactly like in the ordinary $U(1)$ gauge theory. 

Implementing the Gauss constraint the Hamiltonian becomes
\be
H=\sum_{x} \frac{e^2}{2}\Pi_{x,12}^2\;,
\ee
with the constraint \eqref{eq:Tensor_gauss}. We could also derive the Gauss constraint by imposing the gauge invarinace $A_{x,12}\rightarrow A_{x,12}+\Delta_1\Delta_2\phi_x$ on the Hilbert space of the above Hamiltonian directly. The operator which implement this transformation must act as identity on the physical Hilbert space for any choice $\phi_x$, and so
\be
e^{i\sum_x \Delta_1\Delta_2\phi_x \Pi_{x,12}}=\mathbb I\Rightarrow \Delta_1\Delta_2\Pi_{x,12}=0\;,
\ee
In addition we require that $A_{x,12}\rightarrow A_{x,12}+2\pi k_{x,12}$ for any choice of integers $k_{x,12}$. This yields that $\pi_{x,12}$ has an integer spectrum. We can further introduce a $\theta$-term
\be
H=\sum_x \frac{e^2}{2}\left(\Pi_{x,12}-\frac{\theta}{2\pi}\right)^2\;.
\ee
The model is solved by diagonalizing $\Pi_{x,12}$, and the ground state is given as any state of integer eigenvalues $m_{x,12}$ of $\Pi_{x,12}$ which obey the constraint 
\be
\Delta_1\Delta_2 m_{x,12}=0\;.
\ee
The ground state when $-\pi<\theta<\pi$ is simply $m_{x,12}=0$ everywhere, while at $\theta=\pi$, the ground state is given by any configuration $m_{x,12}=c(x_1)$,  or $m_{x,12}=c(x_2)$ where $c(x_{1,2})$ is constrained to be zero or unity. The degeneracy of the ground state is $2^{N_1}+2^{N_2}-2$. 

Note that this model has a large symmetry given by the operator equations
\be
\partial_0\Pi_{x,12}=0
\ee
along with the Gauss law $\Delta_1\Delta_2\Pi_{x,12}=0$. In other words every $\Pi_{x,12}$ is conserved point-wise. This is an exotic 1-form symmetry of the model, where the Gauss law is modified to allow $\Pi_{x,12}$ to be nonconstant, and depend on either only on $x_1$ or only on $x_2$. 

The model allows for a coupling to the scalar field theory we discussed previously. We can write
\be
H=\sum_{x} \frac{1}{2Ja}\pi_x^2+\frac{J}{2a}\left(\Delta_1\Delta_2\phi_x+A_{x,12}+2\pi n_x\right)^2+\sum_x\frac{1}{2\beta a}\left(\Pi_{x,12}-\frac{\theta}{2\pi}\right)^2\;.
\ee
The Gauss law in this case reduces to
\be\label{eq:gauged_XYplaq_gauss}
\Delta_1\Delta_2 \Pi_{x,12}=\pi_{x+\hat 1+\hat 2}
\ee

Alternatively we may choose to couple the gauge fields as an XY-plaquette model instead
\be\label{eq:XYplaq_gauged}
H=\sum_{x} \frac{1}{2Ja}\pi_x^2-\frac{J}{a}\cos\left(\Delta_1\Delta_2\phi+A_{x,12}\right)+\sum_x\frac{e^2}{2\beta a}\left(\Pi_{x,12}-\frac{\theta}{2\pi}\right)^2\;.
\ee

\begin{figure}[htbp] 
   \centering
   \includegraphics[width=4in]{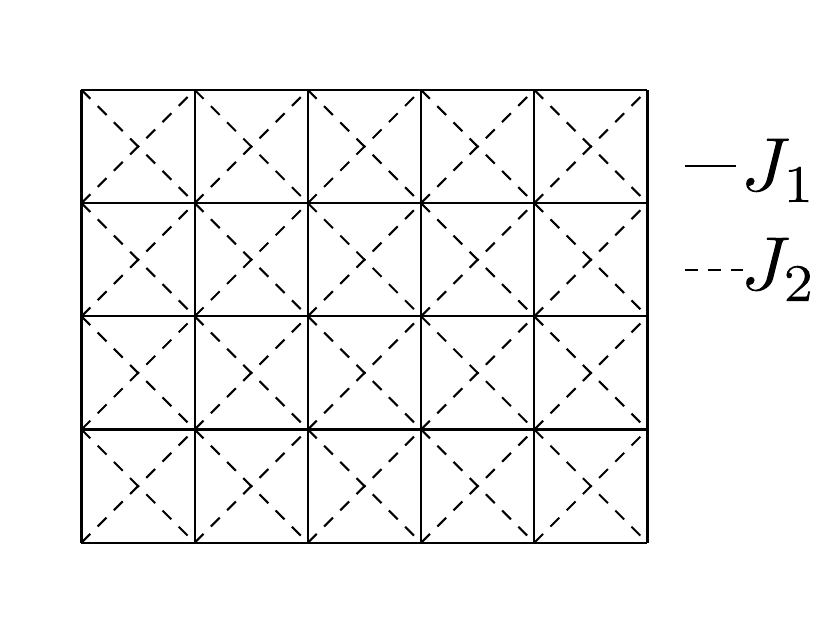} 
   \caption{A schematic depiction of the $J_1-J_2$ 2d Ising model.}
   \label{fig:J1-J2}
\end{figure}

Let us now consider the strong gauge coupling limit $e^2 \rightarrow \infty$ at fixed $a$, and also take $\theta=\pi$. Then $\Pi_{x,12}$ must be $0$ or $1$, as other values have infinite energy. The Hilbert space of the gauge field momentum $\Pi_{x,12}$ gets truncated to only two states, the rest being separated by an infinite energy gap of the order $1/(\beta a)$. We can hence replace $\Pi_{x,12}\rightarrow \frac{\sigma^3_x+1}{2}$, where $\sigma^3_{x}$ is the 3rd Pauli matrix on the site $x$. Moreover we have that $\pi_{x}=\Delta_{1}\Delta_2\Pi_{x-\hat 1-\hat 2,12}\rightarrow \frac{1}{2}\Delta_{1}\Delta_2\sigma^3_{x}$. Finally we can write $\cos(\Delta_1\Delta_2\phi+A_{x,12})= \frac{1}{2}e^{i\Delta_{1}\Delta_2\phi+iA_{x,12}}+\frac{1}{2}e^{-i\Delta_{1}\Delta_2\phi-iA_{x,12}}$. The first of these two changes the eigenvalue of $\Pi_{x,12}$ by $+1$ and the second changes it by $-1$. So we should replace them by $\sigma^+_x$ and $\sigma_{x}^-$ respectively, i.e. we can write
\be
\cos\left(\Delta_1\Delta_2\phi+A_{x,12}\right)\rightarrow \frac{1}{2}(\sigma_{x}^++\sigma_{x}^{-1})=\frac{1}{2}\sigma_x^1\;.
\ee
Finally since $\sum_{x}(\Delta_{1}\Delta_2\sigma_x)^2=2\sum(\sigma_{x}\sigma_{x+\hat 1+\hat 2}-2\sigma_{x}\sigma_{x+\hat 1}-2\sigma_{x}\sigma_{x+\hat 2}+\sigma_{x+\hat 1}\sigma_{x+\hat 2})+\dots$ where the dots indicate an operator proportional to identity, our model reduces to 
\be
H\rightarrow H_{eff}=\sum_{x}\frac{1}{4Ja}\left(\sigma^3_{x}\sigma^3_{x+\hat 1+\hat 2}-2\sigma^3_{x}\sigma^3_{x+\hat 1}-2\sigma^3_{x}\sigma^3_{x+\hat 2}+\sigma^3_{x+\hat 1}\sigma^3_{x+\hat 2}\right)-\sum_{x}\frac{J}{2a}\sigma_{x}^1\;,
\ee
where we dropped the irrelevant constant terms. We can also write the above as
\be
H_{eff}=-J_1\sum_{<xy>}\sigma^3_{x}\sigma_y^3+J_{2}\sum_{<<xy>>}\sigma_{x}^3\sigma_{y}^3-h\sum_{x}\sigma_x^1\;,
\ee
with $J_1=\frac{1}{2aJ}, J_2=J_1/2$ and $h=\frac{J}{2a}$, and where $\sum_{<xy>}$ signifies the sum over next-negboring sites $x$ and $y$, while the $<<x,y>>$ signifies the sum over next-next-neighboring sites (i.e. along diagonals of the square lattice) (see Fig.~\ref{fig:J1-J2}). This model is sometimes called the transverse field $J_1-J_2$ Ising model. The phase diagram of such models has been studied in\footnote{Note that, since the square lattice is bipartite, we can flip the spins on one sublattice and hence effectively flip $J_1\rightarrow -J_1$. Hence the model with both couplings anti-ferromagnetic is equivalent to the $J_1$ ferromagnetic and $J_2$ anti-ferromagnetic.} \cite{kato2015quantum,kellermann2019quantum,oitmaa2020frustrated,sadrzadeh2018phase}.

In particular we are interested in $J_1=2J_2$ case. Let us discuss the $h\rightarrow 0$ limit. In this case the ground state of the model is highly degenerate, as any state which has all the spins along any row (or column) constant is a ground state of the system. This limit corresponds precisely to a $J\rightarrow 0$, which is the free tensor gauge field limit, that also has a degeneracy even at finite gauge coupling. Some degeneracy is guaranteed by the fact that the conserved charge $q_x=\sigma_x^3$ gets flipped under the charge conjugation symmetry $q_x\rightarrow -q_x$. Since the ground states are labeled by some configuration of conserved charges $\{q_x\}$, then a state with $\{-q_x\}$ is also a ground state. Furthermore, since $q_x$ can only be $\pm 1$, we cannot have that $q_x$ and $-q_x$ are equal, and so the two states are distinct. This can be viewed as a mixed anomaly between the symmetry generated by $q_x=\sigma_{x}^3$ and the charge conjugation generated by $C=\sum_x\sigma^1_x$.

How do we understand the huge degeneracy at the point $J_1=2J_2$? Recall that the model arose from the expansion of $(\Delta_1\Delta_2\sigma_x^3)^2$. The ground state needs to minimize this term, which can be thought of as the energetically imposed exotic gauss law \eqref{eq:Tensor_gauss}. But this Gauss law allows a huge number of solutions, rendering the ground state very degenerate. Changing the Gauss law by setting $J_1\ne 2J_2$ will lift a lot of degeneracy, but not all, because of the mixed 't Hooft anomaly between the local symmetry generated by $\sigma_x^3$ and charge conjugation. Indeed if $J_1>2J_2$ the system goes into the striped phase, and when $J_1<2J_2$ it goes into the anti-ferromagnetic N\'eel phase. Both of these break the charge conjugation symmetry and hence are consistent with the 't Hooft anomaly. 

\begin{figure}[t] 
   \centering
   \includegraphics[width=4.5in]{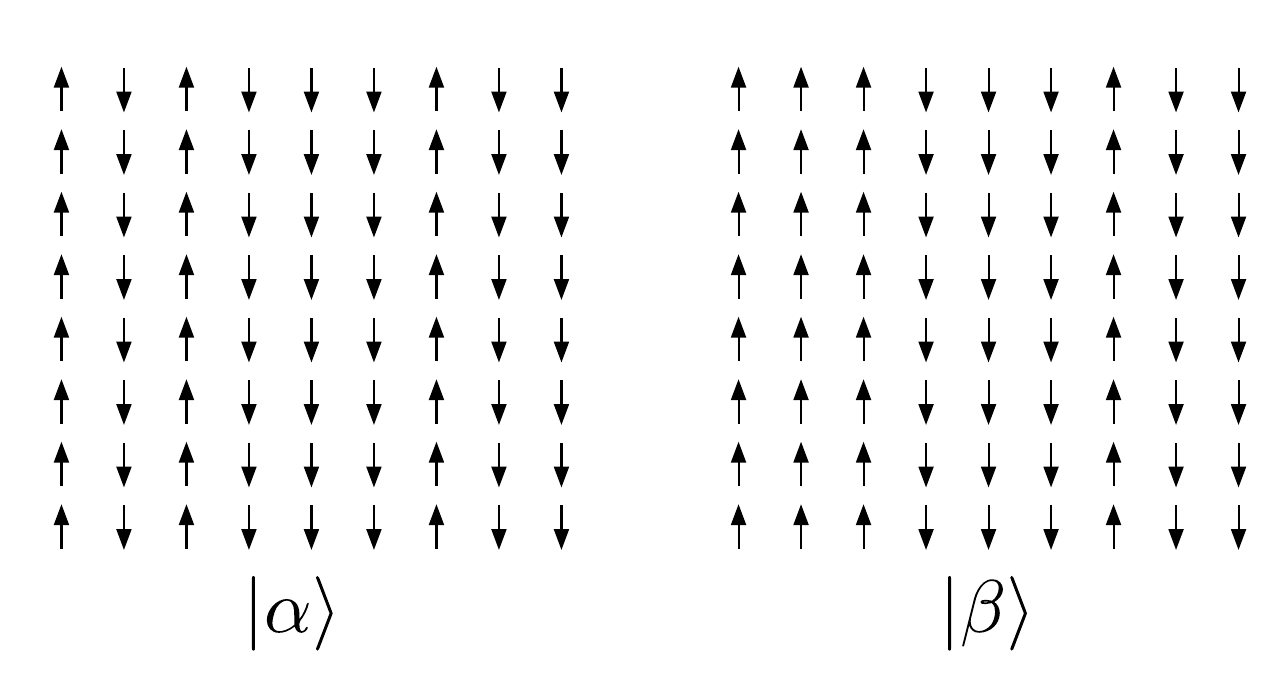} 
   \caption{A graphical depiction of two degenerate states in the $J1=2J_2$ limit of the $h=0$ $J_1-J_2$ Ising model. }
   \label{fig:degeneracy_lift}
\end{figure}

However once $h\ne0$ then $\sigma_x^3$ is no longer conserved, and the reasoning of the above paragraph is violated. A priori there is nothing that prevents the degenerate vacua from lifting. Let us consider two such degenerate states $\ket\alpha$ and $\ket\beta$ at $h=0$, which are depicted in Fig.~\ref{fig:degeneracy_lift}. They differ only by the spins in one of the (i.e. second) columns. If we make the lattice finite, then the leading contribution to the transition probability $\braket{\alpha}{e^{-iHt}}{\beta}$ from one to the other is  $(h/J_1)^{N_{2}}$, where $N_2$ is the number of lattice sites in the $2$-direction of the spatial lattice. Hence in the thermodynamic limit, the two states have no overlap when $h/J_1\ll 1$ and we do not expect degeneracy to be lifted by small fields\footnote{Note that our conclusion is in disagreement with some of the literature \cite{sadrzadeh2016emergence,kellermann2019quantum}.}. If the two degenerate states are even more different and where they differ by $K$ columns, then the splitting is even more suppressed, i.e. by $(h/J_1)^{N_{2}K}$. On the other hand when $h\gg J_1$ we expect a unique ground state polarised in the $\sigma^1_x=1$ direction. A minimal conjecture is then to assume that there is one phase transition and that in the low field phase we have exponential number of degenerate ground states. The nature of the transition is not clear (see \cite{bobak2018frustrated,kellermann2019quantum,sadrzadeh2016emergence,sadrzadeh2018phase}). In \cite{sadrzadeh2016emergence} a transition at the value $h/J_1\approx 0.5$, which translates to $J\approx \sqrt{2}\approx 1.41$. On the other hand in, when $e^2\rightarrow 0$ the model effectively reduces to the ungauged XY-model studied in \cite{paramekanti2002ring}. Unfortunately this work only discussed the XY plaquette model with a chemical potential of the form
\be\label{eq:H_XY-plaquette}
H_{XY-plaquette}=\sum_{x}\left(\frac{U}{2}(\pi_x-\bar n)^2-K\cos(\Delta_1\Delta_2\phi_x)\right)\;,
\ee
where $U,K$ are dimensionful constants and $\mu=U\bar n$ serves as a chemical potential. The reference \cite{paramekanti2002ring} studies a model with $\bar n=1/2$ and finds the transition at $U/K\approx 2.4$. For our gauged model the chemical potential would not do anything, as a finite gauge charge is projected out by the Gauss constraint \eqref{eq:gauged_XYplaq_gauss}. At any rate the gauged XY-plaquette model is expected to have a similar transition at zero gauge coupling $e^2=0$, but potentially of the different nature than the $e^2\ne 0$ transition. This happens in the gauged 1+1d compact scalar, where $e^2=0$ has a BKT transition, while for $e^2\ne0$ an Ising transition is expected \cite{Affleck:1991tj,Sulejmanpasic:2020ubo, Komargodski:2017dmc}.

\begin{figure}[t] 
   \centering
   \includegraphics[width=4.5in]{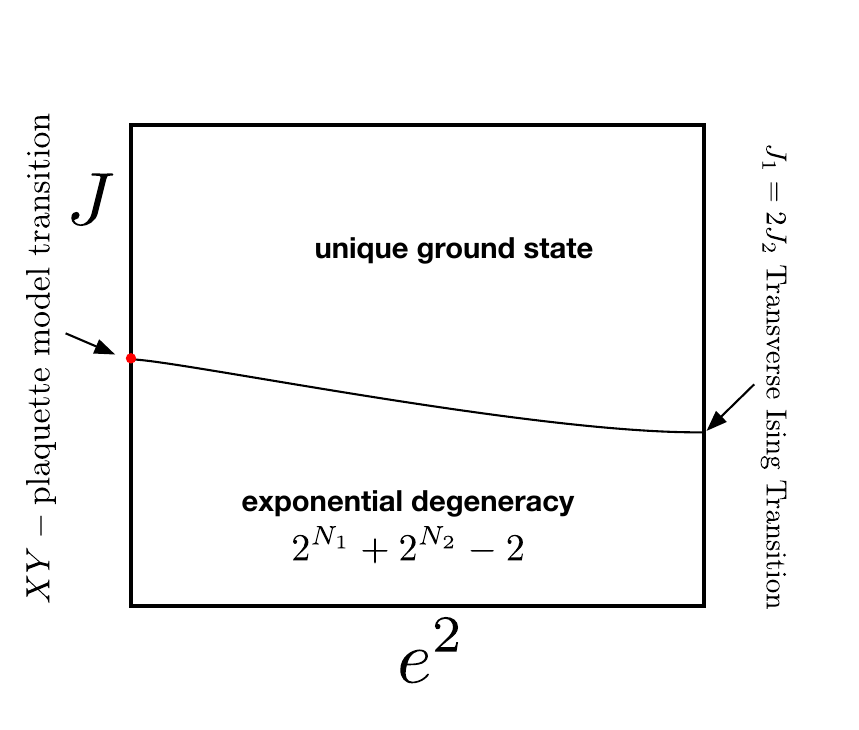} 
   \caption{A phase diagram of the model \eqref{eq:XYplaq_gauged}. The limit $e^2\rightarrow\infty$ is the $J_1-J_2$ Ising model limit, which reportedly has a phase transition at $h/J_1\approx 0.5$, which gives $J\approx \sqrt{2}$. The other extreme should have an ungauged XY-plaquette model transition \eqref{eq:H_XY-plaquette}, which was studied in \cite{paramekanti2002ring} but only at finite chemical potential, where it has a transition for $U/K\approx 2.4$. We conjecture that the nature of the phase transition is the same, save for the limit $e^2=0$.}
   \label{fig:phase_diag}
\end{figure}

We are unaware of numerical studies of the XY-plaquette model with zero chemical potential so we have no way of estimating the $J$ for which the transition is to occur. The phase diagram of our model \eqref{eq:XYplaq_gauged} is shown in Fig.~\ref{fig:phase_diag}. 

\section{Conclusions}
 
In this work we have discussed the construction of Villain Hamiltonians. The construction allows many models to be written down keeping the correct global symmetry and anomaly structures. Moreover, such models reduce to the Modified Villain Action models \cite{Sulejmanpasic:2019ytl} and \cite{Gorantla:2021svj} when the theory is placed on a finite time Euclidean lattice. The Villain Hamiltonian models on the lattice can also be made manifestly self-dual, a feature lacking in both the continuum as well as the Modified Villain Actions. Further, for models which are exactly self-dual, the duality is manifestly a symmetry of the Hamiltonian, although it is embedded into lattice translations in a nontrivial way.

Further, we have shown that coupling the compact scalar models in 1+1d and the exotic fracton compact scalar model in 2+1d to the relevant gauge fields with a $\theta=\pi$ term reduces to the quantum Ising model in a transverse field in 1 and 2 spatial dimensions respectively when the gauge coupling is sent to infinity. This is especially interesting in the case of the gauged XY-plaquette model, where the phase structure of the model could be understood by studying the simpler corresponding Ising model.

The models discussed here can be used to construct Hamiltonian counterparts of models with exact electric magnetic self-duality, which may allow for nontrivial interacting fixed points, like it was done on space-time lattices \cite{Anosova:2022yqx,Sulejmanpasic:2019ytl}, or to construct Hamiltonian versions of the 3d U(1) gauge theories relevant for the search of N\'eel to VBS deconfined criticality \cite{senthil2004deconfined,sandvik2007evidence,sandvik2010continuous,shao2016quantum} which is a yet unsettled question. Villain Hamiltonians may provide a simpler testbeds for the existence of deconfined criticality. On the other hand some bosonic compact scalar models have fermionic duals \cite{Coleman:1974bu,Cao:2022lig} in the continuum, and it is an interesting question whether such duals can be constructed exactly on the lattice, perhaps shedding light into the lattice construction of chiral gauge theories (see \cite{wang2022symmetric,zeng2022symmetric,Tong:2021phe,Razamat:2020kyf} for some recent works on this problem).

\section*{Acknowledgments}

We would like to thank Pavel Buividovich, Tyler Helmuth, Abdoullah Langari, Anders Sandvik, Nathan Seiberg, Shu-Heng Shao, Yuya Tanizaki and David Tong for comments and discussions. This work is supported by the University Research Fellowship of the Royal Society of London. This research is also supported in part by the STFC consolidated grant number ST/T000708/1.

 \begin{appendix}
 
 \section{Solutions to compact scalar models}\label{app:compact_boson}
 
 Here we discuss the solutions of the model \eqref{eq:H1p1d} and \eqref{eq:H1fracton}. We will start with the conventional compact scalar model \eqref{eq:H1p1d} and then discuss the fracton model of \eqref{eq:H1fracton}. Other p-form models can also be solved along similar lines. 
 
 \subsection{Solution to the U(1) scalar in 1 spatial dimension}

We have that the equations of motion coming from the Hamiltonian \eqref{eq:H1p1d} are given by
\begin{align}
&\dot\phi_x=i[H,\phi_x]=\frac{\pi_x}{Ja}\;,\\
&\dot\pi_x=i[H,\pi_x]=\frac{J}{a}\big(\phi_{x+1}-2\phi_x+\phi_{x-1}+2\pi(n_x-n_{x-1})\big)\;,\\
&\label{app:eom3}\dot{\tilde{\phi}}_x=i[H,\tilde\phi_x]=\frac{2\pi J}{a}(\phi_{x+1}-\phi_x+2\pi n_x)\;,\\
&\dot n_x=i[H,n_x]=0\;,\\
&\dot {\tilde n}_x=i[H,\tilde n_x]=0.
\end{align}

The first two equations can be combined to give
\be
\ddot\phi_x=\frac{1}{a^2}\left(\phi_{x+1}-2\phi_x+\phi_{x-1}+2\pi(n_{x}-n_{x-1})\right)\;.
\ee
Now going into momentum space we have
\begin{align}
&\phi_x=\sum_{p} e^{ixp}a_p\;,&&n_x=\sum_p e^{ixp}m_p\;.
\end{align}
where $p$ takes values $p=0,\frac{2\pi}{N},\cdots, \frac{2\pi(N-1)}{N}$.
From the equations of motion we have that $a_p$ obeys
\be\label{eq:ap_eom}
\ddot a_p+{\omega_{p}^2} a_p= \frac{2\pi}{a^2} (1-e^{-i p})m_p\;,
\ee
with $\omega_p=\frac{2|\sin\frac{p}{2}|}{a}$ the constraint $a_p=a_{-p}^\dagger$ and $m_p=m_{-p}^\dagger$. Note also that $\omega_p^2=\frac{1}{a^2}(1-e^{ip})(1-e^{-ip})$. Now note that because of the e.o.m for $n_x$, $m_p$ is constant in time. So we can solve the above equation easily. To do this let us define\footnote{The constant $\sqrt{N\omega_p}$ in front is there for later convenience. } 
\be
b_p(t)=\sqrt{2aNJ\omega_p}\left(a_p(t)-\frac{1-e^{-ip}}{\omega_p^2a^2}2\pi  m_p\right)\;, \qquad p\ne 0
\ee 
We have that the equation of motion in terms of $b_p(t)$ are simply
\be\label{app:bp_eom}
\ddot b_p(t)+{\omega_p^2} b_p(t)=0\;,
\ee
with a Hermitean solution 
\be
b_p(t)=e^{ {i\omega_p} t}B_p^\dagger+e^{ -{i\omega_p} t}B_{-p}\;, \text{ for $p\ne 0$}
\ee
where $B_p$ is a constant operator. For $p=0$ we have from \eqref{eq:ap_eom} that 
\be
a_0=\Phi+\frac{1}{Ja}\Pi t\;.
\ee
where $\Phi$ and $\Pi$ are operators constant in time. We will see later that $\Pi$ is the conjugate momentum to $\Phi$.

Now note that 
\begin{align}
	&\phi_x= \Phi+\frac{1}{Ja}\frac{\Pi t}{N}+\sum_{p\ne0} \sqrt{\frac{1}{2NaJ\omega_p}} \left(B_p^\dagger e^{{i\omega_p} t+ip x}+B_p e^{-{i\omega_p} t-i p x}\right)+\sum_{p\ne0}\frac{e^{i px}}{1-e^{ip}}2\pi m_p\;,\\
	\label{eq:pi_x_exp}
	&\pi_x=\frac{\Pi}{N}+\sum_{p\ne0} \sqrt{\frac{J\omega_p}{2Na}}i\left( B_p^\dagger e^{{i\omega_p} t+i p x}-B_p e^{-{i\omega_p} t-ipx}\right)\;.
\end{align}
We now want to impose canonical commutation relations $[\phi_x,\pi_y]=i\delta_{xy}$. We can take $B_p$ to commute with $m_p$ because $\phi_x$ was taken to commute with $n_x$. So 
\begin{multline}
	[\phi_x,\pi_y]=\frac{1}{N}[\Phi,\Pi]+J\sum_{p\ne 0}\sqrt\frac{\omega_p}{2NaJ}i\left(\frac{1}{N}[\Pi,B_p]e^{-{i\omega_p}t+ipx}-\frac{1}{N}[\Pi,B_p^\dagger]e^{{i\omega_p}t-ipx}\right)\\
	+\sum_{p\ne0} \sqrt{\frac{1}{2NaJ\omega_p}} \left([B_p,\Pi] e^{-{i\omega_p}t+i p x}+[B_p^\dagger,\Pi] e^{{i\omega_p}t-i p x}\right)\\
	+\frac{1}{2Na}\sum_{p,p'\ne0}\sqrt{\frac{\omega_{p'}}{\omega_p }}i\Bigg[[B_p,B_{p'}]e^{i\left(\omega_{p}+\omega_{p'}\right)t+ip x+ip' y}-[B_p^\dagger,B_{p'}^\dagger]e^{-i\left(\omega_p+\omega_{p'}\right)t-i px-i p'y}
	\\ -[B_p,B_{p'}^\dagger]e^{i(\omega_{p}-\omega_{p'})t+i px-i p' y}+[B_p^\dagger ,B_{p'}]e^{i(\omega_{p}-\omega_{p'})t-i px+i p' y}
	\Bigg]=i\delta_{xy}\;,
\end{multline}
where we assumed that $m_p$ commutes with all $B_p$ and $
\Pi$. 
To satisfy the above we must take $[B_p,B_{p'}]=[\Pi,B_p]=0$ (as otherwise the expression would be time-dependent) and $[B_p,B_{p'}^\dagger]=\delta_{p,p'}$, $[\Phi,\Pi]=i$, to reproduce the Kronecker delta. As promised, $\Pi$ is a conjugate momentum to $\Phi$.

Now, note that
\begin{multline}
	\phi_{x+1}-\phi_x+2\pi n_x= \sum_{p } \left((e^{ip}-1)a_p+2\pi m_p\right)e^{i x p}=2\pi m_0+\sum_{p\ne0} \sqrt{\frac{1}{2NJa\omega_p}}(e^{ip}-1)b_p(t)e^{i x p}=\\=2\pi \frac{\tilde\Pi}{N}+\sum_p \sqrt{\frac{1}{2NJa\omega_p}}\left((e^{ip}-1)B_p^\dagger e^{-i\omega_p t+i p x}+(e^{-ip}-1)B_{p} e^{i\omega_p t-i p x}\right)\;,
\end{multline}
where in the last step we identified $m_0=\frac{1}{N}\sum_x n_x=\frac{\tilde \Pi}{N}$, where $\tilde \Pi$ is a ``spatial winding number''\footnote{This idenification comes from defining  $\tilde\Pi=\frac{1}{2\pi}\sum_x (\phi_{x+1}-\phi_x+2\pi n_x)$, which is the lattice variant of $\tilde \Pi=\frac{1}{2\pi}\int dx\partial_x\phi$.}. As we will see, this will also play the role of the momentum operator conjugate to $\tilde\Phi$  -- the zeromode of $\tilde\phi_x$ operator,  
so that the Hamiltonian becomes
\begin{multline}
	H=\frac{1}{2Ja}\sum_x\pi_x^2+\frac{J}{2a}\sum_{x}\left(\phi_{x+1}-\phi_x+2\pi n_x\right)^2=\\=\frac{J(2\pi)^2}{2a}N\left(\frac{\tilde\Pi}{N}\right)^2+\frac{N}{2Ja}\left(\frac{\Pi}{N}\right)^2+\frac{1}{2}\sum_{p\ne0}\omega_{p}\left(B_pB_p^\dagger+B_p^\dagger B_p\right).
\end{multline}

Notice that the equations of motion imply
\be
\dot \pi_x=\frac{1}{2\pi}(\dot{\tilde \phi}_x-\dot{\tilde\phi}_{x-1})\Rightarrow \pi_x=\frac{1}{2\pi}(\tilde\phi_x-\tilde\phi_{x-1}+\hat K_x)\;.
\ee
where $\hat K$ is a constant operator. Now imposing the constraint \eqref{eq:constraint} we must have $\hat K_x=2\pi \tilde n_x$ where $\tilde n_x$ has an integer spectrum. Let us now in analogy to what we done before write
\begin{align}
	&\tilde\phi_x=\sum_{p}\tilde a_pe^{ix p}\;,\\
	&\tilde n_x=\sum_p \tilde m_p e^{i xp}\;.
\end{align}
Then
\begin{multline}
	\tilde \phi_x-\tilde \phi_{x-1}+2\pi \tilde n_x=2\pi \tilde m_0+\sum_{p\ne0}(1-e^{-ip})\underbrace{\left(\tilde a_p+2\pi \frac{\tilde m_p}{1-e^{-ip}}\right)}_{=\frac{2\pi\sqrt{J}}{\sqrt{2Na\omega_p}}\tilde b_p}e^{ixp}=\\=2\pi m_0+{2\pi\sqrt{J}}\sum_{p\ne0}\frac{1-e^{-ip}}{\sqrt{2Na\omega_p}}\tilde b_pe^{ixp}\;.
\end{multline}
Further, e.o.m. also imply
\begin{equation}
	\pi_x= aJ\dot\phi_x=\frac{\Pi}{N}+ J\sum_{p\ne0}\frac{\dot b_p}{\sqrt{2NJa\omega_p}}e^{ixp}\;.
\end{equation}
On the other hand we have by the constraint \eqref{eq:constraint} that
\be
\pi_x=\tilde m_0+\sqrt{J}\sum_{p\ne0} \frac{1-e^{-ip}}{ \sqrt{2Na\omega_p}}\tilde b_pe^{ixp}\;.
\ee
\begin{align}
&\tilde m_0=\frac{\Pi}{N}&&\tilde b_p = \frac{1}{1-e^{-ip}}\dot b_p\;,\qquad p\ne 0\;.
\end{align}
Note that $\sum_{x}\tilde n_x=Nm_0=\Pi$ is the dual-winding charge, which is, of course, the momentum.

Differentiating $\tilde b_p$ relation w.r.t. time twice, we have that
\be
\ddot{\tilde b}_p=-\omega_p^2 \frac{1}{1-e^{-ip}}\dot b_p= -\omega_p^2 \tilde b_p\;,
\ee
where we used the e.o.m.-s \eqref{app:bp_eom} for $b_p$. Hence $\tilde b_p$ also satisfies the harmonic oscillator equations and can be written as
\be
\tilde b_p=\tilde B_p e^{-i\omega_pt}+\tilde B_{-p}^\dagger e^{i\omega_p t}\;.
\ee
Now we have a relation
\be
\tilde B_p= \frac{- i\omega_p}{1-e^{ip}}B_p\;.
\ee
It is easy to check that
\be
[\tilde B_p,\tilde B_{p'}^\dagger]= \delta_{p,p'}\;.
\ee
Finally, we want to show that the winding number $\tilde\Pi=\sum_x n_x$ is the dual momentum. To do this we must show that $\tilde a_0=\tilde\Phi+\frac{(2\pi)^2}{JN}\tilde\Pi t$, where $\tilde\Phi$ is the canonical conjugate to $\tilde \Pi$. Firstly, it is obvious that $\dot{\tilde a}_0=\frac{(2\pi)^2}{JN}\Pi$, from the \eqref{app:eom3}, which is checked by summing that equation w.r.t. $x$. Further, we compute $[\tilde a_0,\tilde\Pi]$ commutator
\be
[\tilde a_0,\tilde\Pi]=\frac{1}{N}\sum_{x,y}\underbrace{[\tilde \phi_x,n_y]}_{i\delta_{x,y}}=i\;,
\ee
hence $a_0=\tilde\Phi+\frac{(2\pi)^2}{JN}\tilde\Pi$.

\subsection{Correlators}\label{app:correlator}

Let us now compute the equal-time correlator $\avg{e^{i\phi_x}e^{-i\phi_y}}$ of the ground state. To do this we will normal order the operators $e^{i\phi_x}$ by putting all the creation operators $B_p^\dagger$ to the left of $B_p$ anihilation operators. Let us write $\phi_x$ as
\be
\phi_x= \phi_x^++\phi_x^-+\phi_x^0\;,
\ee
where
\begin{align}
&\phi_x^0=\Phi+\frac{\Pi t}{JN}+\sum_{p\ne0}\frac{e^{i px}}{1-e^{ip}}2\pi m_p\;,\\
&\phi_x^+=\sum_p \frac{1}{\sqrt{2NaJ\omega_p}} B_p^\dagger e^{i\omega_pt+i px}\;,\\
&\phi_x^-=\sum_p \frac{1}{\sqrt{2NaJ\omega_p}} B_pe^{-i\omega_p t- ipx}\;.
\end{align}
Now let us write
\be
\sum_{p\ne 0}\frac{e^{ipx}}{1-e^{ip}}m_p=\frac{1}{N}\sum_{y} \sum_{p\ne 0} \frac{e^{ip (x-y)}}{1-e^{ip}}n_y\;.
\ee
Since we have that
\be
\sum_{p\ne 0} \frac{e^{i p x}}{1-e^{ip}}=\lim_{\epsilon\rightarrow 0} \sum_{p\ne 0} \frac{e^{ip x}}{1-e^{ip-\epsilon}},
\ee
where the limit $\epsilon\rightarrow 0$ is approached from above, we have that
\be
\frac{1}{1-e^{ip-\epsilon}}=\sum_{s=0}^\infty e^{i sp-s\epsilon}\;.
\ee
Then, since, $\sum_{p\ne0} e^{i p (x+s)}=\sum_{p} e^{ip(x+s)}-1=N\sum_{q\in \mathbb Z}\delta_{x+s,q N}-1 $, we have that
\begin{multline}
\sum_{p\ne 0}\frac{e^{ipx}}{1-e^{ip-\epsilon}}=\sum_{s=0}^\infty \left(N\sum_{q=-\infty}^\infty\delta_{x+s,qN}-1\right)e^{-s\epsilon}=\\=N\sum_{q\ge \frac{x}{N}}e^{-(Nq-x)\epsilon}-\frac{1}{1-e^{-\epsilon}}=\frac{Ne^{\left(\tilde x-N(1-\delta_{\tilde x,0})\right)}}{1-e^{-N\epsilon}}-\frac{1}{1-e^{-\epsilon}},
\end{multline}
where $\tilde x$ is the remainder of the division of $x$ by $N$. So
\be
\sum_{p\ne 0}\frac{e^{ipx}}{1-e^{ip}}=\tilde x-\frac{N+1}{2}\;,
\ee
and hence
\be
\sum_{p\ne 0}\frac{e^{ipx}}{1-e^{ip}}m_p=\frac{1}{N}\sum_z \widetilde{(x-z)}n_z-\left(\frac{1}{2}+\frac{1}{2N}\right)\tilde\Pi\;.
\ee
where we wrote $\tilde \Pi=\sum_x n_x$. 
\be
e^{i\phi_x^0-i\phi_y^0}=e^{\frac{i}{N}{(x-y)}\tilde\Pi}
\ee
Now we look at the expectation value
\be\label{eq:correlator_result}
\avg{:e^{i\phi_x}::e^{-i\phi_y}:}=\avg{e^{i\phi_x^-}e^{-i\phi_y^+}e^{i2\pi \frac{1}{N}(x-y)\tilde\Pi}}=e^{-[\phi_x^-,\phi_y^+]}=e^{-\frac{1}{2JNa}\sum_{p\ne 0} \frac{e^{ip(x-y)}}{\omega_p}}\;,
\ee
where we used the fact that for the ground state $\tilde\Pi=0$.


\subsection{Solution to the 2+1d XY-plaquette compact scalar fracton model}\label{app:XY_sol}
Here we discuss the Hamiltonian \eqref{eq:H1fracton}. The equations of motion are given by
\begin{align}
&\dot\phi_x=i[H,\phi_x]=\frac{1}{aJ}\pi_x\;,\\
&\dot\pi_x=i[H,\pi_x]=-\frac{J}{a}\left(\Delta_1^2\Delta_2^2\phi_{x-\hat 1-\hat 2}+2\pi \Delta_1\Delta_2 n_{x-\hat 1-\hat 2}\right)\;,\\
&\dot\varphi_x=i[H,\varphi_x]=-\frac{2\pi J}{a}(\Delta_1\Delta_2\phi_x+2\pi n_x)\;,\\
&\dot n_x=0.
\end{align}
We proceed similarly to the case of compact scalar in 2d. We write
\begin{align}
&\phi_x=\sum_p a_pe^{ixp}\;,\\
&n_x=\sum_p q_p e^{ixp}\;,
\end{align}
and, by combining the e.o.m. for $\dot\phi_x$ and $\dot\pi_x$ we get 
\be
\ddot\phi_x+\frac{1}{a^2}(\Delta_1^2\Delta_2^2\phi_{x-\hat 1-\hat 2}+2\pi \Delta_{1}\Delta_2 n_{x-\hat 1-\hat 2})=0
\ee
from where it follows that
\be
\ddot a_p+\omega_p^2 a_p=-\frac{2\pi q_p}{a^2} (1-e^{-ip_1})(1-e^{-ip_2})\;.
\ee
where $\omega_p=\frac{4}{a}|\sin\frac{p_1}{2}||\sin\frac{p_2}{2}|$. 
When neither $p_1$ nor $p_2$ are zero we can define
\be
b_p= c_p\left(a_p+\frac{2\pi}{(1-e^{ip_1})(1-e^{ip_2})}q_p\right)\;,
\ee
where $c_p$ are some constants and $b_p$ now satisfies the equation
\be
\ddot b_p+\omega_p^2 b_p=0\;,
\ee
with the general solution
\be
b_p=B_p^\dagger e^{i\omega_pt}+B_{-p} e^{-i\omega_pt}\;.
\ee
On the other hand, when either $p_1$ or $p_2$ is zero, we have that the e.o.m. for $\phi_x$ is either purely a function of $x_1$ or purely a function of $x_2$. We therefore get
\begin{align}
&\phi_x=\Phi_0+\Phi_1(x_1)+\Phi_2(x_2)+\frac{-\Pi_0/(N_1N_2)+\Pi_1(x_1)/N_2+\Pi_2(x_2)/N_1}{aJ}t\nonumber\\
&\qquad+\sum_{p_1\ne0\;, p_2\ne 0}  \frac{1}{c_p}\left(B_{p}^\dagger e^{i\omega_p t+ix\cdot p }+B_{p} e^{-i\omega_p t-ix\cdot p}\right)-\sum_{p_1\ne0\;,p_2\ne 0}\frac{2\pi e^{ix\cdot p}}{(1-e^{ip_1})(1-e^{p_2})}q_p\\
&\label{eq:pi_x_plqt}\pi_x=\frac{-\Pi_0}{N_1N_2}+\frac{\Pi_1(x_1)}{N_2}+\frac{\Pi_2(x_2)}{N_1}+ \sum_{p_1\ne0\;, p_2\ne 0}\frac{Ja\omega_pi}{c_p} \left(B_{p}^\dagger e^{i\omega_p t+ix\cdot p }-B_{p}e^{-i\omega_p t-ix\cdot p}\right)\;.
\end{align}
Note that we have captured the zero modes by three pieces: a piece only dependent on $x_1$, only dependent on $x_2$ and a constant piece. This is redundant, as the constant piece is already captured by the pieces which depend on $x_1$ and $x_2$, but it will be convenient.
Imposing the commutation relation $[\phi_x,\pi_y]=i\delta_{x,y}$ is equivalent to demanding that
\begin{align}
&[\Phi_1(x_1),\Pi_1(y_1)]=i\delta_{x_1,y_1}\;,&& [\Phi_2(x_2),\Pi_2(y_2)]=i\delta_{x_2,y_2}\;,\\
&[\Phi_0,\Pi_0]=i\;, &&[B_p,B_{p'}^\dagger]=\frac{c_p^2}{2N_1N_1aJ\omega_p}\delta_{p,p'}\qquad \;,
\end{align}
with all other commutator combinations being zero.
If we set $c_p=\sqrt{2N_1N_1Ja\omega_p}$ the last commutator simplifies to $[B_p,B_{p'}^\dagger]=\delta_{p,p'}$. As we noted before, the decomposition into $\Phi_0,\Phi_1(x_1)$ and $\Phi_2(x_2)$ is ambiguous, because we could shift these operators as follows
\begin{equation}
\begin{array}{l}
\Phi_0\rightarrow \Phi_0+\delta\;,\\
\Phi_1(x_1)\rightarrow \Phi_1(x_1)+\delta_1\;,\\
\Phi_2(x_1)\rightarrow \Phi_2(x_1)+\delta_2\;,
\end{array}\qquad \text{such that $\delta+\delta_1+\delta_2=0$}\;,
\end{equation}
where $\delta,\delta_1$ and $\delta_2$ are constants. 
The above invariance enforces a constraint
\be
\sum_{x_1}\Pi_1(x_1)=\sum_{x_2}\Pi_2(x_2)=\Pi_0\;.
\ee
Further we also can shift 
\begin{equation}
\begin{array}{l}
\Pi_0\rightarrow \Pi_0+N_1N_2\Lambda\;,\\
\Phi_1(x_1)\rightarrow \Phi_1(x_1)+N_2\Lambda_1\;,\\
\Phi_2(x_1)\rightarrow \Phi_2(x_1)+N_1\Lambda_2\;,
\end{array}\qquad \text{such that $-\Lambda+\Lambda_1+\Lambda_2=0$}\;.
\end{equation}
which enforces a constraint
\be
N_2\sum_{x_1}\Phi_1(x_1)=N_1\sum_{x_2}\Phi_2(x_2)=-N_1N_2\Phi_0.
\ee

Now let us write 
\begin{multline}
\Delta_1\Delta_2 \phi_{x}=\sum_{p_1\ne0\;, p_2\ne 0}\frac{1}{\sqrt{2N_1N_2 Ja\omega_p}}\\
\times\Bigg[(e^{i p_1}-1)(e^{ip_2}-1)B_pe^{ip\cdot x+i\omega_p t}+(e^{-i p_1}-1)(e^{-ip_2}-1) B_p^\dagger e^{-ip\cdot x-i\omega_pt}\Bigg]\\-2\pi \sum_{p_1\ne0\;, p_2\ne 0}e^{i x\cdot p}q_p\;.
\end{multline}
Now we write
\be
\sum_{p_1\ne0\;, p_2\ne 0}q_pe^{ix\cdot p}=n_x-\frac{1}{N_1}\sum_{x_1}n_x-\frac{1}{N_2}\sum_{x_2}n_x+\frac{1}{N_1N_2}\sum_{x}n_x\;,
\ee
so the Hamiltonian is given by
\begin{multline}
H=\frac{1}{2JaN_1^2N_2^2}\sum_{x}\Big({\Pi_0}-N_1\Pi_1(x_1)-N_2\Pi_2(x_2)\Big)^2\\+\frac{J}{2aN_1^2N_2^2}\sum_{x}\left(\tilde\Pi_0-N_1\tilde\Pi_1(x_1)-N_2\tilde\Pi_2(x_2)\right)^2\\+\sum_{p} \omega_p\left(B_p^\dagger B_p+\frac{1}{2}\right)\;.
\end{multline}
We can simultaneously diagonalize $\Pi_0,\Pi_{1,2}(x_{1,2)}$ and their tilde counter-parts, along with $B_p^\dagger B_p$, to obtain the spectrum.

The model also has a tensor symmetry. The symmetry current is given by
\begin{align}
&J_{0,x}=\pi_x \;, &&J^{12}=-\frac{J}{a}\left(\Delta_1\Delta_2\phi_{x-\hat 1-\hat 2}+2\pi n_{x-\hat 1-\hat 2}\right)\;.
\end{align}
We have that
\be
\partial_0J_x^0-\Delta_1\Delta_2J^{12}_x=0\;,
\ee
by the equations of motion, which means that charges
\begin{align}
&Q_1(x_1)=\sum_{x_2}J_{0,x}\;,\\
&Q_2(x_2)=\sum_{x_1}J_{0,x}
\end{align}
are conserved. Indeed since we have $\sum_{x_i}J_{0,x}=\sum_{x_i}\pi_x=\Pi_i(x_i)$ as can be easily checked by plugging $\pi_x$ from equation \eqref{eq:pi_x_plqt} and using the fact that $\sum_{x_i}\Pi_i(x_i)=\Pi_0$.

\section{Linking number}\label{app:linking}

Consider an Euclidean manifold $M_D$ of dimension $D$, two submanifold of $M_D$, $\Sigma_{D-p}$ and $\Sigma'_{p+1}$ of dimensions $D-p$ and $p+1$, respectively. We will take that $\Sigma_{D-p}$ and $\Sigma'_{p+1}$ have a boundary which are, respectively, $D-p-1$ and $p$ dimensional. We want to define the linking number of the boundaries $\partial\Sigma_{D-p}$ and $\partial\Sigma'_{p+1}$. 

\begin{figure}[htbp] 
   \centering
   \includegraphics[width=4in]{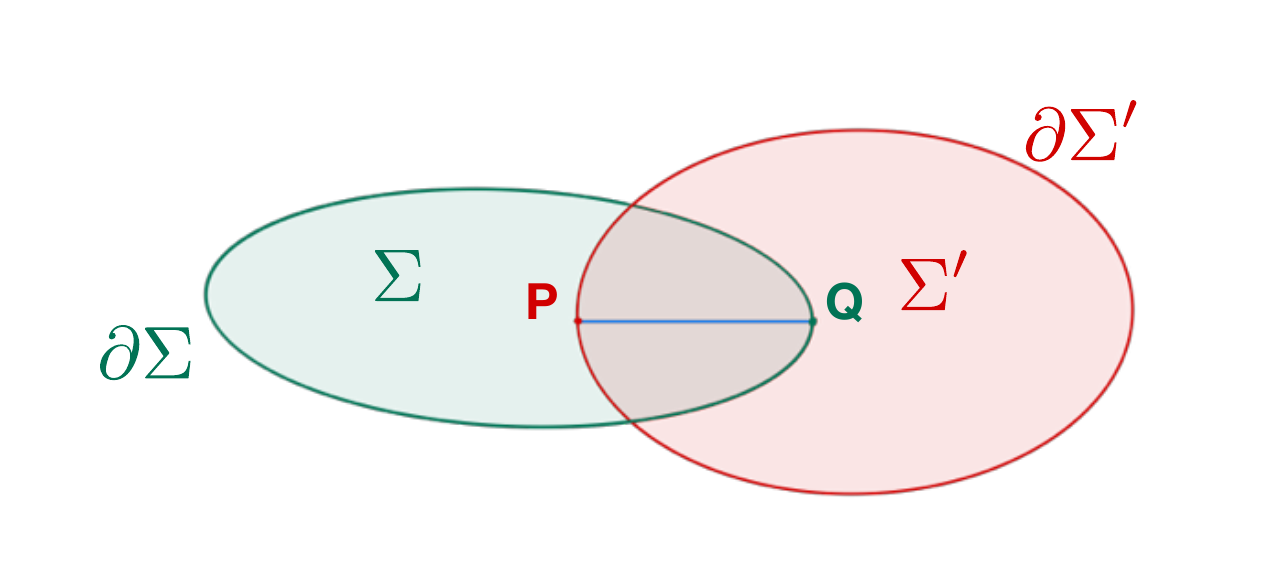} 
   \caption{Sketch of the intersection between the two submanifolds of $M_D$, $\Sigma_{D-p}$ and $\Sigma'_{p+1}$.}
   \label{fig:Sigma_Sigma_prime}
\end{figure}

We sketch the situation in Fig.~\ref{fig:Sigma_Sigma_prime}. Let $X^\mu_{\Sigma}(\sigma^1,\sigma^2,\dots,\sigma^{D-p})$ be the local coordinates in $M_D$ describing $\Sigma$, and $\sigma^i, i=1,\dots, D-p$ are parameters parametrizing $\Sigma$ (i.e. world-volume coordinates). Similarly we have $X^\mu_{\Sigma'}(\sigma^{'1},\dots,\sigma^{'p+1})$ describing $\Sigma'$. Now let us choose world-volume coordinates such that $\sigma_i=0$ is the point $P$ on $\Sigma$ where $\partial\Sigma'$ pierces $\Sigma$, and $\sigma_i'=0$ is the point $Q$ on $\Sigma'$ where the boundary of $\partial\Sigma$ pierces $\Sigma'$. Further, we will take that the line where $\Sigma$ and $\Sigma'$ intersect is described by $X^\mu_{\Sigma}(\sigma^1,0,\dots,0)$ and $X^\mu_{\Sigma}(\sigma^{'1},0,\dots,0)$, where $X_\Sigma^\mu(0,\dots,0)$ and $X_{\Sigma'}^\mu(1,0,\dots,0)$ describe the point $P$  and $X^\mu_{\Sigma'}(0,\dots,0)$ and $X^\mu_{\Sigma}(1,0,\dots,0)$ describe the point $Q$.

Now we define the linking number of the boundaries $\partial \Sigma$ and $\partial\Sigma'$ as the number of times that $\partial\Sigma'$ intersects $\Sigma$, where we take the sign of the contribution to be determined as follows. If $\partial\Sigma'$ intersects $\Sigma$ in such a way that the product of their tangent spaces $(T\Sigma)_P\times (T\partial\Sigma')_P$ at point $P$, has the same orientation as the tangent space of $(TM)_P$, then we will take point $P$ to contribute with a positive sign to the linking number. So we define
\be
L(\partial \Sigma,\partial\Sigma')= I(\Sigma,\partial\Sigma')
\ee
where $I(\Sigma,\partial\Sigma')$ is the net intersection number between $\partial\Sigma'$ and $\Sigma$ in the sense described above.

Let us now show that 
\be
L(\partial\Sigma,\partial\Sigma')=(-1)^{(D-p-1)(p-1)}L(\partial\Sigma',\partial\Sigma)\;.
\ee
To do this we consider the tangent space  $T\Sigma\Big|_P$. It is given by the bases
\be
\left(\frac{\partial}{\partial\sigma^1},\frac{\partial}{\partial\sigma^2},\dots\frac{\partial}{\partial\sigma^{D-p}}\right).
\ee
On the other hand the tangent space  $T\partial\Sigma'\Big|_P$ is given by the basis
\be
\left(\frac{\partial}{\partial\sigma^{'2}},\frac{\partial}{\partial\sigma^{'3}},\dots\frac{\partial}{\partial\sigma^{'p+1}}\right)\;.
\ee
The $(T\Sigma)_P\times (T\partial\Sigma')_{P}$ is given by the basis
\be\label{eq:basis1}
(\frac{\partial}{\partial\sigma^1},\frac{\partial}{\partial\sigma^2},\dots\frac{\partial}{\partial\sigma^{D-p}},\frac{\partial}{\partial\sigma^{'2}},\frac{\partial}{\partial\sigma^{'3}},\dots\frac{\partial}{\partial\sigma^{'p+1}})\;.
\ee
On the other hand we have that the tangent space of $(T\Sigma')_{Q}\times (T\partial\Sigma)_{Q}$ is given by 
\be
(\frac{\partial}{\partial\sigma^{'1}},\frac{\partial}{\partial\sigma^{'2}},\dots\frac{\partial}{\partial\sigma^{'p+1}},\frac{\partial}{\partial\sigma^{2}},\frac{\partial}{\partial\sigma^{3}},\dots\frac{\partial}{\partial\sigma^{D-p}})\;.
\ee
In fact both of these tangent spaces are well defined on the curve joining the two points $P$ and $Q$. On this curve we have that the vector $\frac{\partial}{\partial\sigma^1}$ is equal to $-\frac{\partial}{\partial\sigma^{'1}}$, so we can write the above as
\be\label{eq:basis2}
(-\frac{\partial}{\partial\sigma^{1}},\frac{\partial}{\partial\sigma^{'2}},\dots\frac{\partial}{\partial\sigma^{'p+1}},\frac{\partial}{\partial\sigma^{2}},\frac{\partial}{\partial\sigma^{3}},\dots\frac{\partial}{\partial\sigma^{D-p}})\;.
\ee
Now the basis above on the curve connecting $P$ and $Q$ differs from \eqref{eq:basis1} by a sign\footnote{We need to push all primed vectors to the right in \eqref{eq:basis2}, which gives $(-1)^{p(D-p-1)}=(-1)^{Dp}$. In addition, since the first vector of \eqref{eq:basis1} and $\eqref{eq:basis2}$ differ by a sign, the net contribution is $(-1)^{Dp+1}$.} $(-1)^{Dp+1}$.

 \end{appendix}

\bibliographystyle{JHEP}
\bibliography{bibliography.bib}

\providecommand{\href}[2]{#2}\begingroup\raggedright\begin{thebibliography}{10}

\bibitem{Luscher:1998pqa}
M.~Luscher, \emph{{Exact chiral symmetry on the lattice and the Ginsparg-Wilson
  relation}}, \href{https://doi.org/10.1016/S0370-2693(98)00423-7}{\emph{Phys.
  Lett. B} {\bfseries 428} (1998) 342}
  [\href{https://arxiv.org/abs/hep-lat/9802011}{{\ttfamily hep-lat/9802011}}].

\bibitem{Ginsparg:1981bj}
P.H.~Ginsparg and K.G.~Wilson, \emph{{A Remnant of Chiral Symmetry on the
  Lattice}}, \href{https://doi.org/10.1103/PhysRevD.25.2649}{\emph{Phys. Rev.
  D} {\bfseries 25} (1982) 2649}.

\bibitem{Neuberger:1997fp}
H.~Neuberger, \emph{{Exactly massless quarks on the lattice}},
  \href{https://doi.org/10.1016/S0370-2693(97)01368-3}{\emph{Phys. Lett. B}
  {\bfseries 417} (1998) 141}
  [\href{https://arxiv.org/abs/hep-lat/9707022}{{\ttfamily hep-lat/9707022}}].

\bibitem{Sulejmanpasic:2019ytl}
T.~Sulejmanpasic and C.~Gattringer, \emph{{Abelian gauge theories on the
  lattice: $\theta$-Terms and compact gauge theory with(out) monopoles}},
  \href{https://doi.org/10.1016/j.nuclphysb.2019.114616}{\emph{Nucl. Phys. B}
  {\bfseries 943} (2019) 114616}
  [\href{https://arxiv.org/abs/1901.02637}{{\ttfamily 1901.02637}}].

\bibitem{Gorantla:2021svj}
P.~Gorantla, H.T.~Lam, N.~Seiberg and S.-H.~Shao, \emph{{A modified Villain
  formulation of fractons and other exotic theories}},
  \href{https://doi.org/10.1063/5.0060808}{\emph{J. Math. Phys.} {\bfseries 62}
  (2021) 102301} [\href{https://arxiv.org/abs/2103.01257}{{\ttfamily
  2103.01257}}].

\bibitem{Choi:2021kmx}
Y.~Choi, C.~Cordova, P.-S.~Hsin, H.T.~Lam and S.-H.~Shao, \emph{{Noninvertible
  duality defects in 3+1 dimensions}},
  \href{https://doi.org/10.1103/PhysRevD.105.125016}{\emph{Phys. Rev. D}
  {\bfseries 105} (2022) 125016}
  [\href{https://arxiv.org/abs/2111.01139}{{\ttfamily 2111.01139}}].

\bibitem{Cheng:2022sgb}
M.~Cheng and N.~Seiberg, \emph{{Lieb-Schultz-Mattis, Luttinger, and 't Hooft --
  anomaly matching in lattice systems}},
  \href{https://arxiv.org/abs/2211.12543}{{\ttfamily 2211.12543}}.

\bibitem{Yoneda:2022qpj}
M.~Yoneda, \emph{{Equivalence of the modified Villain formulation and the dual
  Hamiltonian method in the duality of the XY-plaquette model}},
  \href{https://arxiv.org/abs/2211.01632}{{\ttfamily 2211.01632}}.

\bibitem{Anosova:2022yqx}
M.~Anosova, C.~Gattringer, N.~Iqbal and T.~Sulejmanpasic, \emph{{Phase
  structure of self-dual lattice gauge theories in 4d}},
  \href{https://doi.org/10.1007/JHEP06(2022)149}{\emph{JHEP} {\bfseries 06}
  (2022) 149} [\href{https://arxiv.org/abs/2203.14774}{{\ttfamily
  2203.14774}}].

\bibitem{Anber:2018jdf}
M.M.~Anber and E.~Poppitz, \emph{{Anomaly matching, (axial) Schwinger models,
  and high-T super Yang-Mills domain walls}},
  \href{https://doi.org/10.1007/JHEP09(2018)076}{\emph{JHEP} {\bfseries 09}
  (2018) 076} [\href{https://arxiv.org/abs/1807.00093}{{\ttfamily
  1807.00093}}].

\bibitem{Misumi:2019dwq}
T.~Misumi, Y.~Tanizaki and M.~\"Unsal, \emph{{Fractional $\theta$ angle, 't
  Hooft anomaly, and quantum instantons in charge-$q$ multi-flavor Schwinger
  model}}, \href{https://doi.org/10.1007/JHEP07(2019)018}{\emph{JHEP}
  {\bfseries 07} (2019) 018}
  [\href{https://arxiv.org/abs/1905.05781}{{\ttfamily 1905.05781}}].

\bibitem{Cherman:2022ecu}
A.~Cherman, T.~Jacobson, M.~Shifman, M.~Unsal and A.~Vainshtein,
  \emph{{Four-fermion deformations of the massless Schwinger model and
  confinement}},  \href{https://arxiv.org/abs/2203.13156}{{\ttfamily
  2203.13156}}.

\bibitem{Komargodski:2017dmc}
Z.~Komargodski, A.~Sharon, R.~Thorngren and X.~Zhou, \emph{{Comments on Abelian
  Higgs Models and Persistent Order}},
  \href{https://doi.org/10.21468/SciPostPhys.6.1.003}{\emph{SciPost Phys.}
  {\bfseries 6} (2019) 003} [\href{https://arxiv.org/abs/1705.04786}{{\ttfamily
  1705.04786}}].

\bibitem{Komargodski:2017smk}
Z.~Komargodski, T.~Sulejmanpasic and M.~\"Unsal, \emph{{Walls, anomalies, and
  deconfinement in quantum antiferromagnets}},
  \href{https://doi.org/10.1103/PhysRevB.97.054418}{\emph{Phys. Rev. B}
  {\bfseries 97} (2018) 054418}
  [\href{https://arxiv.org/abs/1706.05731}{{\ttfamily 1706.05731}}].

\bibitem{Kikuchi:2017pcp}
Y.~Kikuchi and Y.~Tanizaki, \emph{{Global inconsistency, \textquoteright{}t
  Hooft anomaly, and level crossing in quantum mechanics}},
  \href{https://doi.org/10.1093/ptep/ptx148}{\emph{PTEP} {\bfseries 2017}
  (2017) 113B05} [\href{https://arxiv.org/abs/1708.01962}{{\ttfamily
  1708.01962}}].

\bibitem{Sulejmanpasic:2020ubo}
T.~Sulejmanpasic, \emph{{Ising model as a $U(1)$ lattice gauge theory with a
  $\theta$-term}},
  \href{https://doi.org/10.1103/PhysRevD.103.034512}{\emph{Phys. Rev. D}
  {\bfseries 103} (2021) 034512}
  [\href{https://arxiv.org/abs/2009.13383}{{\ttfamily 2009.13383}}].

\bibitem{paramekanti2002ring}
A.~Paramekanti, L.~Balents and M.P.~Fisher, \emph{Ring exchange, the exciton
  bose liquid, and bosonization in two dimensions}, {\emph{Physical Review B}
  {\bfseries 66} (2002) 054526}.

\bibitem{Seiberg:2020bhn}
N.~Seiberg and S.-H.~Shao, \emph{{Exotic Symmetries, Duality, and Fractons in
  2+1-Dimensional Quantum Field Theory}},
  \href{https://doi.org/10.21468/SciPostPhys.10.2.027}{\emph{SciPost Phys.}
  {\bfseries 10} (2021) 027}
  [\href{https://arxiv.org/abs/2003.10466}{{\ttfamily 2003.10466}}].

\bibitem{Seiberg:2020wsg}
N.~Seiberg and S.-H.~Shao, \emph{{Exotic $U(1)$ Symmetries, Duality, and
  Fractons in 3+1-Dimensional Quantum Field Theory}},
  \href{https://doi.org/10.21468/SciPostPhys.9.4.046}{\emph{SciPost Phys.}
  {\bfseries 9} (2020) 046} [\href{https://arxiv.org/abs/2004.00015}{{\ttfamily
  2004.00015}}].

\bibitem{Seiberg:2020cxy}
N.~Seiberg and S.-H.~Shao, \emph{{Exotic $\mathbb{Z}_N$ symmetries, duality,
  and fractons in 3+1-dimensional quantum field theory}},
  \href{https://doi.org/10.21468/SciPostPhys.10.1.003}{\emph{SciPost Phys.}
  {\bfseries 10} (2021) 003}
  [\href{https://arxiv.org/abs/2004.06115}{{\ttfamily 2004.06115}}].

\bibitem{Gorantla:2020xap}
P.~Gorantla, H.T.~Lam, N.~Seiberg and S.-H.~Shao, \emph{{More Exotic Field
  Theories in 3+1 Dimensions}},
  \href{https://doi.org/10.21468/SciPostPhys.9.5.073}{\emph{SciPost Phys.}
  {\bfseries 9} (2020) 073} [\href{https://arxiv.org/abs/2007.04904}{{\ttfamily
  2007.04904}}].

\bibitem{ma2018higher}
H.~Ma and M.~Pretko, \emph{Higher-rank deconfined quantum criticality at the
  lifshitz transition and the exciton bose condensate}, {\emph{Physical Review
  B} {\bfseries 98} (2018) 125105}.

\bibitem{Gorantla:2020jpy}
P.~Gorantla, H.T.~Lam, N.~Seiberg and S.-H.~Shao, \emph{{fcc lattice,
  checkerboards, fractons, and quantum field theory}},
  \href{https://doi.org/10.1103/PhysRevB.103.205116}{\emph{Phys. Rev. B}
  {\bfseries 103} (2021) 205116}
  [\href{https://arxiv.org/abs/2010.16414}{{\ttfamily 2010.16414}}].

\bibitem{Gorantla:2021bda}
P.~Gorantla, H.T.~Lam, N.~Seiberg and S.-H.~Shao, \emph{{Low-energy limit of
  some exotic lattice theories and UV/IR mixing}},
  \href{https://doi.org/10.1103/PhysRevB.104.235116}{\emph{Phys. Rev. B}
  {\bfseries 104} (2021) 235116}
  [\href{https://arxiv.org/abs/2108.00020}{{\ttfamily 2108.00020}}].

\bibitem{Gorantla:2022eem}
P.~Gorantla, H.T.~Lam, N.~Seiberg and S.-H.~Shao, \emph{{Global dipole
  symmetry, compact Lifshitz theory, tensor gauge theory, and fractons}},
  \href{https://doi.org/10.1103/PhysRevB.106.045112}{\emph{Phys. Rev. B}
  {\bfseries 106} (2022) 045112}
  [\href{https://arxiv.org/abs/2201.10589}{{\ttfamily 2201.10589}}].

\bibitem{Gorantla:2022ssr}
P.~Gorantla, H.T.~Lam, N.~Seiberg and S.-H.~Shao, \emph{{2+1d Compact Lifshitz
  Theory, Tensor Gauge Theory, and Fractons}},
  \href{https://arxiv.org/abs/2209.10030}{{\ttfamily 2209.10030}}.

\bibitem{Burnell:2021reh}
F.J.~Burnell, T.~Devakul, P.~Gorantla, H.T.~Lam and S.-H.~Shao, \emph{{Anomaly
  inflow for subsystem symmetries}},
  \href{https://doi.org/10.1103/PhysRevB.106.085113}{\emph{Phys. Rev. B}
  {\bfseries 106} (2022) 085113}
  [\href{https://arxiv.org/abs/2110.09529}{{\ttfamily 2110.09529}}].

\bibitem{distler2022spontaneously}
J.~Distler, A.~Karch and A.~Raz, \emph{Spontaneously broken subsystem
  symmetries}, {\emph{Journal of High Energy Physics} {\bfseries 2022} (2022)
  1}.

\bibitem{dirac2001lectures}
P.A.M.~Dirac, \emph{Lectures on quantum mechanics}, vol.~2, Courier Corporation
  (2001).

\bibitem{kato2015quantum}
Y.~Kato and T.~Misawa, \emph{Quantum tricriticality in antiferromagnetic ising
  model with transverse field: a quantum monte carlo study}, {\emph{Physical
  Review B} {\bfseries 92} (2015) 174419}.

\bibitem{kellermann2019quantum}
N.~Kellermann, M.~Schmidt and F.~Zimmer, \emph{Quantum ising model on the
  frustrated square lattice}, {\emph{Physical Review E} {\bfseries 99} (2019)
  012134}.

\bibitem{oitmaa2020frustrated}
J.~Oitmaa, \emph{Frustrated transverse-field ising model}, {\emph{Journal of
  Physics A: Mathematical and Theoretical} {\bfseries 53} (2020) 085001}.

\bibitem{sadrzadeh2018phase}
M.~Sadrzadeh and A.~Langari, \emph{Phase diagram of the frustrated j1- j2
  transverse field ising model on the square lattice},  in \emph{Journal of
  Physics: Conference Series}, vol.~969, p.~012114, IOP Publishing, 2018.

\bibitem{sadrzadeh2016emergence}
M.~Sadrzadeh, R.~Haghshenas, S.~Jahromi and A.~Langari, \emph{Emergence of
  string valence-bond-solid state in the frustrated j 1- j 2 transverse field
  ising model on the square lattice}, {\emph{Physical Review B} {\bfseries 94}
  (2016) 214419}.

\bibitem{bobak2018frustrated}
A.~Bob{\'a}k, E.~Jur{\v{c}}i{\v{s}}inov{\'a}, M.~Jur{\v{c}}i{\v{s}}in and
  M.~{\v{Z}}ukovi{\v{c}}, \emph{Frustrated spin-1 2 ising antiferromagnet on a
  square lattice in a transverse field}, {\emph{Physical Review E} {\bfseries
  97} (2018) 022124}.

\bibitem{Affleck:1991tj}
I.~Affleck, \emph{{Nonlinear sigma model at Theta = pi: Euclidean lattice
  formulation and solid-on-solid models}},
  \href{https://doi.org/10.1103/PhysRevLett.66.2429}{\emph{Phys. Rev. Lett.}
  {\bfseries 66} (1991) 2429}.

\bibitem{senthil2004deconfined}
T.~Senthil, A.~Vishwanath, L.~Balents, S.~Sachdev and M.P.~Fisher,
  \emph{Deconfined quantum critical points}, {\emph{Science} {\bfseries 303}
  (2004) 1490}.

\bibitem{sandvik2007evidence}
A.W.~Sandvik, \emph{Evidence for deconfined quantum criticality in a
  two-dimensional heisenberg model with four-spin interactions},
  {\emph{Physical review letters} {\bfseries 98} (2007) 227202}.

\bibitem{sandvik2010continuous}
A.W.~Sandvik, \emph{Continuous quantum phase transition between an
  antiferromagnet and a valence-bond solid in two dimensions: Evidence for
  logarithmic corrections to scaling}, {\emph{Physical review letters}
  {\bfseries 104} (2010) 177201}.

\bibitem{shao2016quantum}
H.~Shao, W.~Guo and A.W.~Sandvik, \emph{Quantum criticality with two length
  scales}, {\emph{Science} {\bfseries 352} (2016) 213}.

\bibitem{Coleman:1974bu}
S.R.~Coleman, \emph{{The Quantum Sine-Gordon Equation as the Massive Thirring
  Model}}, \href{https://doi.org/10.1103/PhysRevD.11.2088}{\emph{Phys. Rev. D}
  {\bfseries 11} (1975) 2088}.

\bibitem{Cao:2022lig}
W.~Cao, M.~Yamazaki and Y.~Zheng, \emph{{Boson-fermion duality with subsystem
  symmetry}}, \href{https://doi.org/10.1103/PhysRevB.106.075150}{\emph{Phys.
  Rev. B} {\bfseries 106} (2022) 075150}
  [\href{https://arxiv.org/abs/2206.02727}{{\ttfamily 2206.02727}}].

\bibitem{wang2022symmetric}
J.~Wang and Y.-Z.~You, \emph{Symmetric mass generation}, {\emph{Symmetry}
  {\bfseries 14} (2022) 1475}.

\bibitem{zeng2022symmetric}
M.~Zeng, Z.~Zhu, J.~Wang and Y.-Z.~You, \emph{Symmetric mass generation in the
  1+ 1 dimensional chiral fermion 3-4-5-0 model}, {\emph{Physical Review
  Letters} {\bfseries 128} (2022) 185301}.

\bibitem{Tong:2021phe}
D.~Tong, \emph{{Comments on symmetric mass generation in 2d and 4d}},
  \href{https://doi.org/10.1007/JHEP07(2022)001}{\emph{JHEP} {\bfseries 07}
  (2022) 001} [\href{https://arxiv.org/abs/2104.03997}{{\ttfamily
  2104.03997}}].

\bibitem{Razamat:2020kyf}
S.S.~Razamat and D.~Tong, \emph{{Gapped Chiral Fermions}},
  \href{https://doi.org/10.1103/PhysRevX.11.011063}{\emph{Phys. Rev. X}
  {\bfseries 11} (2021) 011063}
  [\href{https://arxiv.org/abs/2009.05037}{{\ttfamily 2009.05037}}].

\end{thebibliography}\endgroup

\end{document}